\def\kms{\hbox{km s$^{-1}$}}
\def\hi{H{\sc i}\,}
\def\hii{H{\sc ii}\,}
\def\coa{\hbox{CO(2-1)}}
\def\cob{\hbox{$^{13}$CO(2-1)}}
\def\coc{\hbox{C$^{18}$O(2-1)}}
\def\hcn{\hbox{HCN(3-2)}}
\def\hco{\hbox{HCO$^+$(3-2)}}
\def\msun{M$_{\odot}$\,}
\def\jyb{Jy beam$^{-1}$}
\def\mjyb{mJy beam$^{-1}$}

\def\cm2{cm$^{-2}$}
\def\cm3{cm$^{-3}$}

\def\radec{\hbox{RA, Dec. (J2000)}}

\def\gra{$^{\circ}$}

\def\h{$^{\mathrm h}$}
\def\m{$^{\mathrm m}$}
\def\s{$^{\mathrm s}$}
\documentclass{aa}  

\usepackage{graphicx}
\usepackage{txfonts}
\usepackage{xcolor}
\usepackage{soul}
\usepackage{float}
\begin{document}

   \title{ Physical characterization of S169: A prototypical IR bubble associated with the  massive star-forming region  IRAS\,12326-6245   }

   \titlerunning{Physical characterization of the IR bubble S169}

   \author{N. U. Duronea\thanks{Member of
    the Carrera del Investigador Cient\'{\i}fico of CONICET,
    Argentina.}\inst{1}
          \and
          S. Cichowolski$^{\star}$\inst{2}
          \and 
          L. Bronfman\inst{3}
          \and
          E. Mendoza\inst{4}
          \and
          R.  Finger\inst{3}
          \and
          L. A. Suad$^{\star}$ \inst{2}
          \and
          M. Corti$^{\star}$\inst{5,6}
          \and 
          E. M. Reynoso$^{\star}$\inst{2}
          }

   \institute{Instituto de Astrof\'{\i}sica de La Plata (UNLP - CONICET), La Plata, Argentina.
              \email{nduronea@fcaglp.unlp.edu.ar}
         \and
              Instituto de Astronom\'{\i}a y F\'{\i}sica del Espacio (UBA, CONICET), CC 67, Suc. 28, 1428 Buenos Aires, Argentina 
         \and 
             Departamento de Astronom{\'{\i}}a, Universidad de Chile, Casilla 36, Santiago de Chile
         \and     
             Observatório do Valongo, Universidade Federal do Rio de Janeiro, Ladeira Pedro Antônio, 43, Rio de Janeiro - RJ, 20080-090, Brazil
          \and
             Facultad de Ciencias Astronómicas y Geofísicas, Universidad Nacional de La Plata, Paseo del Bosque s/n, 1900 La Plata, Argentina
          \and  Instituto Argentino de Radioastronom\'{\i}a (CCT-La Plata, CONICET; CICPBA), C.C. No. 5, 1894,Villa Elisa, Argentina     
              }

   \date{Received July 31; accepted December 4}

 
  \abstract
   {   }
   {  With the aim of studying the physical properties of Galactic IR bubbles and to explore their impact  in massive star formation, we present a study of the IR bubble S169, associated with the massive star forming region IRAS\,12326-6245.  }
   { We used \coa,\cob, \coc, \hcn, and \hco\  line data obtained  with
the  APEX telescope using the on-the-fly full sampling technique to study the properties of the molecular gas in the nebula and the IRAS source . To analyze the properties and distribution of the dust, we made use of images obtained from the IRAC-GLIMPSE,  {\it Herschel}, and ATLASGAL archives. The properties of the ionized gas in the nebula were studied using radio continuum and H$\alpha$ images obtained from the SUMSS survey and SuperCOSMOS database, respectively. In our search for  stellar and protostellar objects in the region, we used point source calalogs obtained from the MSX, WISE, GLIMPSE, 2MASS, AAVSO, ASCC-2.5V3, and GAIA databases. }
   {The new APEX observations allowed us to identify three molecular components, each one associated with different regions of the nebula, namely:\ at $-$39 \kms (component A), $-$25 \kms (component B), and $-$17 \kms (component C). Component A is shown to be the  most dense and clumpy. Six molecular condensations (MC1 to MC6) were identified in this component, with MC3 (the densest and more massive one) being  the molecular counterpart of IRAS\,12326-6245.   For this source, we estimated an  H$_2$ column density up to 8$\times$10$^{23}$ cm$^{-2}$.   An LTE analysis of the high density tracer lines \hco\ and \hcn\ on this source, assuming 50 and 150 K, respectively, indicates column densities of $N$(HCO$^+$) = (5.2 $\pm$ 0.1) $\times$ 10$^{13}$~cm$^{-2}$ and $N$(HCN) = (1.9 $\pm$ 0.5) $\times$ 10$^{14}$~cm$^{-2}$.     To explain the morphology and velocity of components A, B, and C , we propose a simple model consisting of a partially complete semisphere-like structure expanding at $\sim$ 12 \kms.    The introduction of this model has led  to a discussion about the distance to both S169 and  IRAS\,12326-6245, which was estimated to be  $\sim$ 2 kpc.     Several candidate YSOs were identified,  projected mostly onto the molecular condensations MC3, MC4, and MC5, which indicates that the star-formation process is very active at the borders of the nebula.   A comparison between observable and modeled parameters was not enough to discern whether  the  collect-and-collapse mechanism is acting at the edge of S169. However, other  processes such as radiative-driven implosion  or even a combination of both mechanisms, namely, collect-and-collapse and radiative-driven implosion,  could be acting simultaneously in the region.}
   {}


   \maketitle
%

\section{Introduction}\label{intro}

Massive stars modify the interstellar medium (ISM),  where they are born through their stellar winds and UV photons, creating interstellar bubbles and \hii regions. These nebulae are detected in the optical and radio continuum ranges, showing the presence of ionized gas, and at infrared (IR), millimeter,  and submillimeter wavelengths, indicating the presence of both dust (at different temperatures) and molecules.

Milky Way surveys at IR wavelengths  have provided  a plethora of examples of bubbles/\hii regions,   where their physical properties can be  obtained on pc/sub-pc scales. Making use of the 8.0 $\mu$m {\it Spitzer}-GLIMPSE survey,   \citet{ben03}       and \citet{chur06,chur07} identified more than 600 candidates for interstellar IR dust bubbles between Galactic longitudes of $-$60\gra \ to 60\gra.  Follow-up observations of the stellar and prestellar populations around some of these bubbles have begun to link them with sites of recent triggered massive star formation (e.g., \citealt{zav06,wat08,deh08,deh09,sam14,ken16,dur17}). These kinds of studies provide important insights into the evolution of interstellar bubbles, the characteristics of the interstellar medium where they evolve, the role of massive stars on favoring or suppressing new generations of stars, and the physical conditions under which massive  stars may be induced to form. 
\begin{figure*}
   \centering
   \includegraphics[width=14cm]{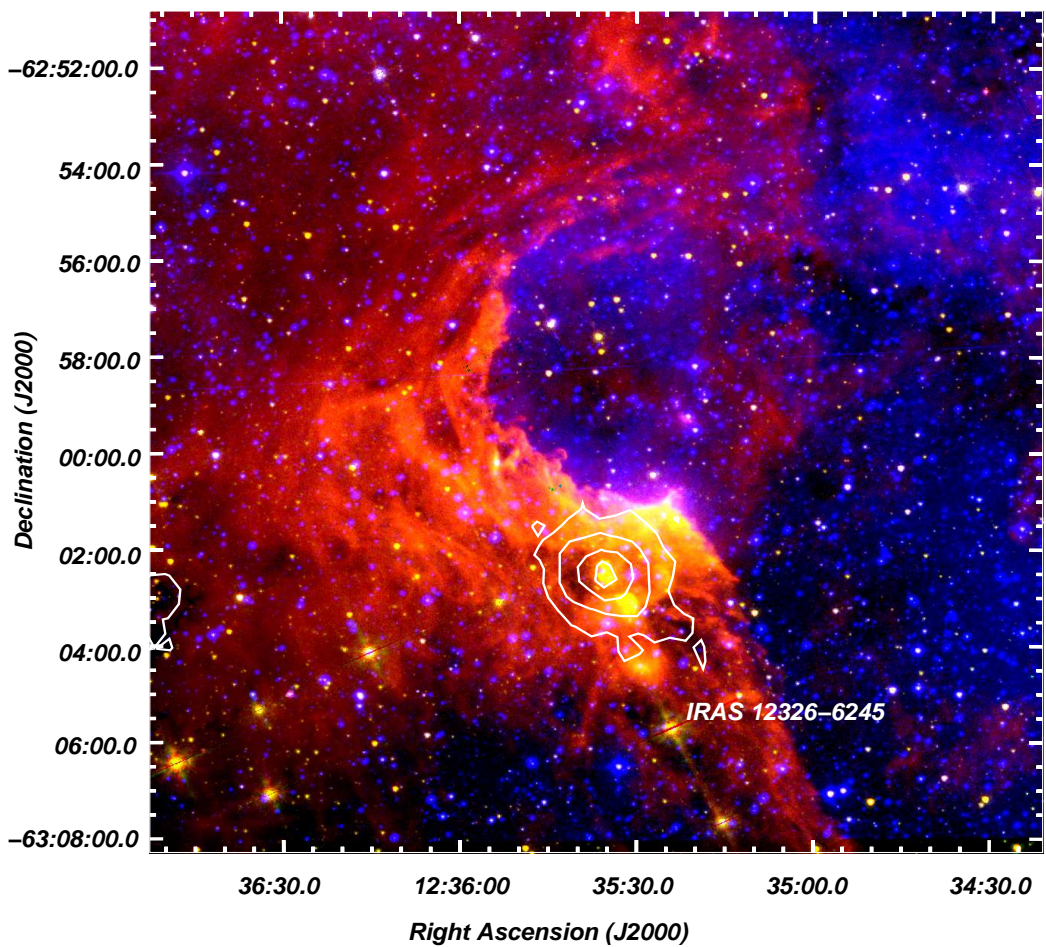}
      \caption{Composite image of S169 and its environs showing  the (IRAC-GLIMPSE) emissions at 8.0 (red) and 5.8 $\mu$m (green).  The superCOSMOS H${\alpha}$ emission is in blue.   White contours depict the continuum emission at 870 $\mu$m (ATLASGAL) arising from IRAS 12326-6245.}
         \label{irac-sc}
   \end{figure*}
As  part  of  a  project  aimed  at  characterizing and studying  the  physical  properties  of  Galactic IR  bubbles  and their surroundings, and  to  better  understand  their  influence in massive  star  formation,  we selected a southern field containing S169 \citep{chur06}, which is a prototypical IR bubble-shaped nebula of about 8 arcmin in diameter,  centered at \radec =  (12\h 35\m 29\fs 2, $-$62\gra58\arcmin03\farcs8).
Figure~\ref{irac-sc} shows a composite image of S169 in 8.0  and 5.8 $\mu$m (IRAC-GLIMPSE) and the H${\alpha}$ (superCOSMOS) emissions. The presence of emission in the 8-$\mu$m band, mostly dominated by the emission of polycyclic aromatic hydrocarbons molecules  (PAHs) at 7.7 and 8.6 $\mu$m, indicates the existence of a  photodissociation region (PDR) probably created by a nearby stellar source.  Since PAHs molecules  are destroyed inside the ionized gas of an \hii\ region \citep{pov07}, but fluorescing when irradiated with weak ultraviolet radiation, they indicate the limits of the ionization front and delineate the boundaries of the bubble nebula, tracing the distribution of the parental molecular gas in which S169 is  developing.  This scenario is also supported by the presence of H${\alpha}$ emission at the center of the bubble nebula.  A distinctive feature in the emission of S169 is the presence of two bright borders at the southern region of the IR nebula, which are discernible in Fig.~\ref{irac-sc} in yellow tonalities.   The spatial distribution of the IR emission  gives a good  hint of the distribution of the molecular gas and the location of the densest regions, very likely associated with the two bright borders described before. As can be seen from Fig.\ref{irac-sc}, the IR emission at 8.0 $\mu$m toward the west of the cavity is  much weaker and composed of a faint arc-like filament approximately at  RA =  12\h 35\m 10\s.

 The nebula S169  is placed  near the IR source IRAS\,12326-6245,  which has characteristics of a young stellar object (YSO), according to the photometric criteria from \citet{jun92}. Substantial evidence suggests that IRAS\,12326-6245 is a dense molecular core  where massive star formation is taking place, namely:  {\it a)} this source was observed in the CS(2-1) line at a velocity of $-$39.4 \kms\   \citep{bro96,ost97}. It was also observed by \citet{zin00} in several lines of HNCO at frequencies between   290 and 806 GHz. This indicates undoubtedly   the existence of dense molecular gas associated with the source; {\it b)} emission from dense dust has been previously detected at 1.2mm as a compact source \citep{fau04,hill05,mie06}. The source is also a strong emitter at 870 $\mu$m     (\citealt{ded11,ko17}; see Fig.~\ref{irac-sc}); {\it c)} CH$_3$OH  maser emission,    a common phenomenon in massive star forming regions,   was reported  by \citet{cas09}  at a velocity of $-$39.8 \kms. OH and H$_2$O maser emission was also detected at the source \citep{cas98,mcleo98}; {\it d)} a variety of complex organic molecules have been detected in the source \citep{ara05,ded11}. These kinds of molecules are  mostly formed in the surface of dust grains and later evaporate into the gas phase. This is  characteristic of ``hot molecular cores'' (HMC; \citealt{kur00}), which are better known to be associated with early stages  of massive star formation;  {\it e)} molecular line maps indicate the presence of a very powerful bipolar outflow at the center of the source \citep{hen00}.

In this work, we present a complete multifrequency  analysis of the ISM around S169 using high spatial resolution molecular  observations, which are necessary to establish the excitation conditions in the molecular environs, as well as to estimate their physical properties and to investigate the kinematics of the region. The analysis of the molecular observations  are complemented with an analysis of a  multifrequency archival dataset, which altogether contribute to our understanding of the impact of stellar feedback and how the ionized gas interacts with its molecular environment.  We also investigate the status of the star-formation processes in the region. In that context, the proximity between S169 and  IRAS\,12326-6245 is crucial for the analysis since it will allow us to better understand the high-mass star-formation process around IR dust bubbles. An analysis of dense molecular line tracers and submillimeter emission in  IRAS\,12326-6245  is also  presented. 

The source IRAS\,12326-6245 is believed to be placed at a distance of 4.4 kpc. Such distance was first estimated by  \citet{ost97} and consistently adopted  afterwards by several authors  (e.g., \citealt{hen00,fau04,ded11}). The distance to IRAS\,12326-6245 is also matter of some debate in this work.


\section{Observations and databases}\label{obs}

\subsection{Molecular observations}

The molecular observations presented in this paper were made in December 2016 with the Atacama Pathfinder EXperiment\footnote{This publication is based on data acquired with the Atacama Pathfinder Experiment (APEX) under program ID  C-098.F-9801. APEX is a collaboration between the Max-Planck-Institut fur Radioastronomie, the European Southern Observatory, and the Onsala Space Observatory} (APEX) telescope \citep{gus06} at Llano de Chajnantor (Chilean Andes). As the front end for the observations, we used the APEX-1 receiver of the  Swedish Heterodyne Facility Instrument (SHeFI;  \citealt{vas08}). The back end for all observations was the eXtended bandwidth Fast Fourier Transform Spectrometer2 (XFFTS2),  with a 2.5 GHz bandwidth divided into 32768 channels. The observed transitions and basic observational parameters are summarized in Table \ref{lines}. Calibration was done by the chopper-wheel technique and the output intensity scale given by the system is $T_{\rm A}$, which represents the antenna temperature corrected for atmospheric attenuation.  The observed intensities were converted to the main-beam brightness temperature scale by $T_{\rm mb}$ = $T_{\rm A}$/$\eta_{\rm mb}$, where   $\eta_{\rm mb}$ is the main beam efficiency. For the APEX-1 receiver we adopted $\eta_{\rm mb}$ = 0.75.

 Observations were made using the on-the-fly (OTF) mode with a step size of 10$^{\prime\prime}$ (full sampling) in two orthogonal scan directions along RA and Dec.(J2000). For the lines \coa, \cob, \ and \coc \ the observed region was  of $\sim$14$'\times$14$'$ in size  centered on \radec = (12\h 35\m 44\s, $-$63\gra00\arcmin08\arcsec). For the lines \hcn\ and \hco\ the observed region was   $\sim$4$'\times$4$'$ in size  centered on the position of the source  IRAS 12326-6245 at \radec = (12\h 35\m 33\s, $-$63\gra02\arcmin56\arcsec). The data reduction was carried out with the CLASS90 package of the IRAM GILDAS software\footnote{http://www.iram.fr/IRAMFR/GILDAS}. 

\begin{table}
\begin{center}
\caption{Observational parameters for the observed transitions.   }
\label{lines}
\begin{tabular}{lcccr}
\hline
molecular   & Frequency &  Beam & Velocity & rms  \\
transition      &  (GHz)    &  ($''$)     & resolution & noise \\
      &       &     & (\kms) & (K)\\ 

\hline
\hline
\coa      & 230.538000         &  $\sim$27       &  0.3      &   $\sim$0.3 \\
\cob     &  220.398677         &  $\sim$28     &   0.3     &  $\sim$0.3  \\
\coc     &  219.560357         &  $\sim$28      & 0.3       &  $\sim$0.25 \\
\hcn     &  265.886180         &  $\sim$23    &  0.3      &  $\sim$0.15\\
\hco     &  267.557526         &  $\sim$23   &  0.3        &  $\sim$0.15    \\
\hline 
\end{tabular}
\end{center}
\end{table}

\subsection{Archival data}

 Infrared data:  {\it a)} Images of ATLASGAL at 870 $\mu$m (345 GHz)  \citep{sch09}. This survey covers the inner Galactic plane, $l$ = 300\gra\ to 60\gra, $|b|$ $\leq$ 1.\gra5, with the root mean square (rms) noise in the range of 0.05 - 0.07 \jyb. The beam size at 870 $\mu$m is 19\farcs2.  {\it b)} Images from the  {\it Herschel}\footnote{ {\it Herschel} is an ESA space observatory with science instruments  provided by European-led Principal Investigator consortia and with important participation from NASA (http://www.cosmos.esa.int/web/herschel/science-archive)} Infrared GALactic (Hi-GAL) plane survey key program \citep{mol10}. We used images from  the photometric array camera and spectrometer (PACS) survey at 70 and 100 $\mu$m, with FWHM of 5\farcs5 and 11\arcsec, respectively, and from the  spectral and photometric imaging receiver (SPIRE) at 350~$\mu$m with a FHWM of 25\arcsec.   {\it c)} Images from {\it Spitzer} at 5.8 and  8.0 $\mu$m from the Galactic Legacy Infrared Mid-Plane Survey Extraordinaire ({\it Spitzer}-GLIMPSE, \citealt{ben03}), retrieved from the Spitzer Science Center\footnote{http://scs.spitzer.caltech.edu}. The images have a spatial resolution of $\sim$ 2\arcsec.\\ 

\noindent Radio continuum data: Mosaics obtained from the Sydney University Molonglo Sky Survey\footnote{http://www.astrop.physics.usyd.edu.au/mosaics}    \citep[SUMSS; ][]{sumss}. This widefield radio imaging survey covers the southern sky at 843 MHz. The resolution is $43^{\prime\prime} \times 43^{\prime\prime}cosec(\delta)$ and the rms noise level is $\sim 1$ mJy\, beam$^{-1}$. \\

\noindent \hi \ data: Data cubes from the Southern Galactic Plane Survey \citep[SGPS; ][]{sgps}.  This survey includes the regions limited by Galactic longitude $253^\circ \leq l \leq 358^\circ$ (SGPS-1) and $5^\circ \leq l \leq 20^\circ$ (SGPS-2) and Galactic latitude $|b| \leq 1$\hbox{$.\!\!{}^\circ$}5. Interferometric observations obtained with the Australia Telescope Compact Array (ATCA) are combined with single dish data from the 64 m Parkes radiotelescope. The final products have an angular resolution of $\sim$ 2\arcmin\  and a rms sensitivity of $\sim 1$ mJy\, beam$^{-1}$.\\

\noindent Narrow band H$\alpha$ data: Retrieved from the SuperCOSMOS H$\alpha$ Survey\footnote{http://www-wfau.roe.ac.uk/sss/halpha/hapixel.html} (SHS). The images have a spatial resolution of $\sim$1$''$  \citep{p05}.\\

\noindent To investigate ionizing stars candidates and YSO candidates in the region of the nebula, we made use of the IRAS Point Source Catalog\footnote{https://irsa.ipac.caltech.edu/Missions/iras.html} \citep{bei88}, the MSX Infrared Point Source Catalog\footnote{https://irsa.ipac.caltech.edu/Missions/msx.html} \citep{ega03}, the  WISE All-Sky Source Catalog\footnote{https://irsa.ipac.caltech.edu/Missions/wise.html} \citep{wri10}, the GLIMPSE point source catalog\footnote{https://irsa.ipac.caltech.edu/Missions/spitzer.html}  \citep{ben03}, the Two Micron All Sky Survey (2MASS)\footnote{https://irsa.ipac.caltech.edu/Missions/2mass.html}  \citep{Skru06}, the American Association of Variable Star Observers (AAVSO) Photometric All-Sky SurveyA (APASS)\footnote{https://www.aavso.org/apass} \citep{hen10}, the ASCC-2.5 V3: All-sky Compiled Catalog of 2.5 million stars \citep{kar09}, and the Astrometric catalog GAIA Data Release 2 \footnote{https://www.cosmos.esa.int/web/gaia/home}  (DR2; \citet{GAIA16, GAIA18}). \\

\begin{figure}
   \centering
   \includegraphics[width=8cm]{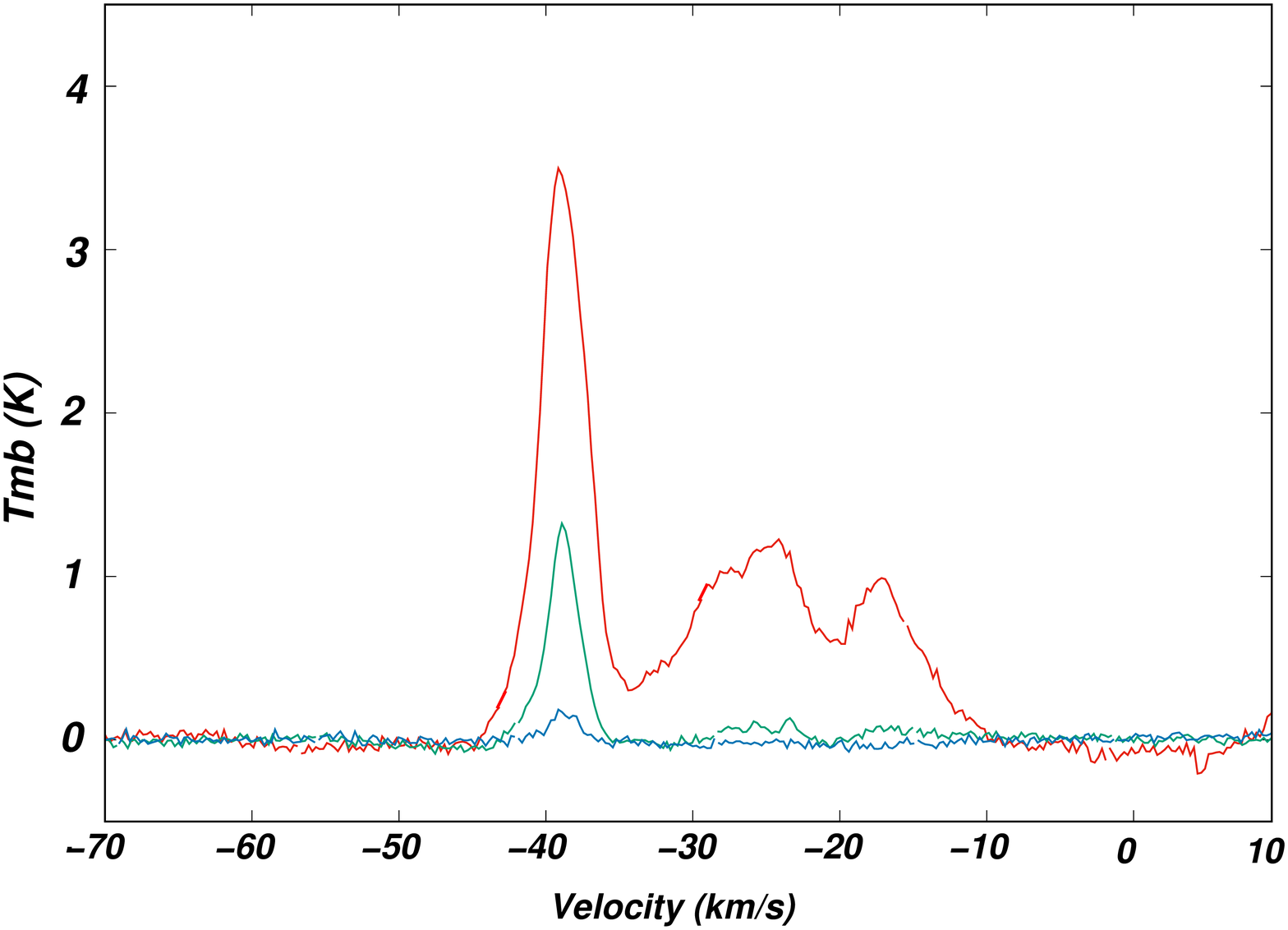}
      \caption{ Total averaged spectra of \coa\ (red), \cob\  (green), and \coc\  (blue)  obtained within a region of  $\sim$14$'\times$14$'$     centered on  \radec = (12\h 35\m 44\s, $-$63\gra00\arcmin08\arcsec).  }
         \label{espectrotot}
   \end{figure}
   
   \begin{figure*}
   \centering
   \includegraphics[width=17cm]{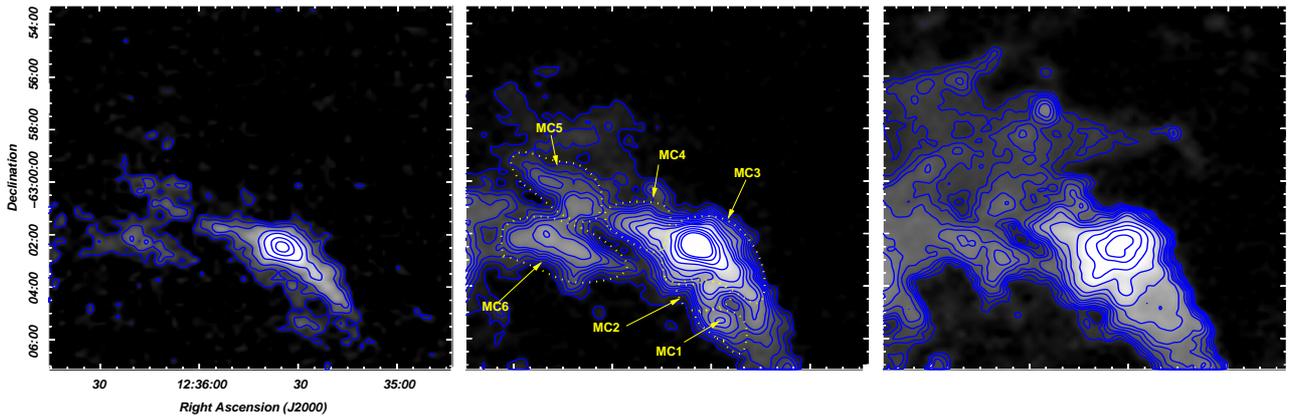}
      \caption{ Spatial distribution of the C$^{18}$O(2-1) emission (left panel), $^{13}$CO(2-1) emission (middle panel), and CO(2-1) emission (right panel) in the velocity interval from $\sim$ --42 \kms\ to --37\, \kms\ (component A). In the left panel, the contours are 0.35 ($\sim$ 3 rms), 0.7, 1.15, 2, 3, 5, and 7 K \kms. In the middle panel the contour levels are 0.5 ($\sim$ 6 rms), 1, 1.5, 2, 2.5, 3, 4, 5, 7, 9, 11, 13, and 15 K \kms. In the right panel the contour levels are 1.5 ($\sim$ 20 rms), 2, 2.5, 3, 4, 5, 6, 8, 10, 14, 18, 22, and 26 K \kms\ ($T_{\rm mb}$).  Identified molecular condensations are indicated in yellow over the \cob\ emission.}
         \label{coprom}
   \end{figure*}

\section{Molecular and dust emission}\label{molecydust}

\subsection{Emission of CO and its isotopologs}\label{coisot}

 Figure~\ref{espectrotot} portrays the total averaged spectra of \coa, \cob, and \coc\  in  a region of  $\sim$14$'\times$14$'$ in a position that is approximately at the center of the bubble nebula. The CO emission clearly depicts a strong narrow velocity component peaking at $\sim$ --39 \kms\, and two weaker broad velocity components peaking at $\sim$ --25 \kms\ and  $\sim$ --17 \kms. For the sake of the analysis, these components will be hereafter referred to as component A, component B, and component C, respectively.  
 
 The peak velocity of component A,  which is very likely dominated by the emission of IRAS\,12326-6245,  is almost coincident with the peak velocity of reported non-masing and masing emission  lines towards the IRAS source \citep{bro96,zin00,hen00,ded11,ara05,cas98,cas09,mcleo98}. This component is also detected in $^{13}$CO and C$^{18}$O emissions, which indicates that it is composed of high density molecular gas. Components B and C, on the other hand, are barely detected in the  $^{13}$CO emission and not detected in the  C$^{18}$O emission. 
 As opposed to  the case of the velocity of component A, neither masing nor non-masing emission lines  were reported for IRAS\,12326-6245 at the velocity of components B and C.

\subsection{Component A}\label{compa}   
  
\begin{figure*}
   \centering
   \includegraphics[width=16.6cm]{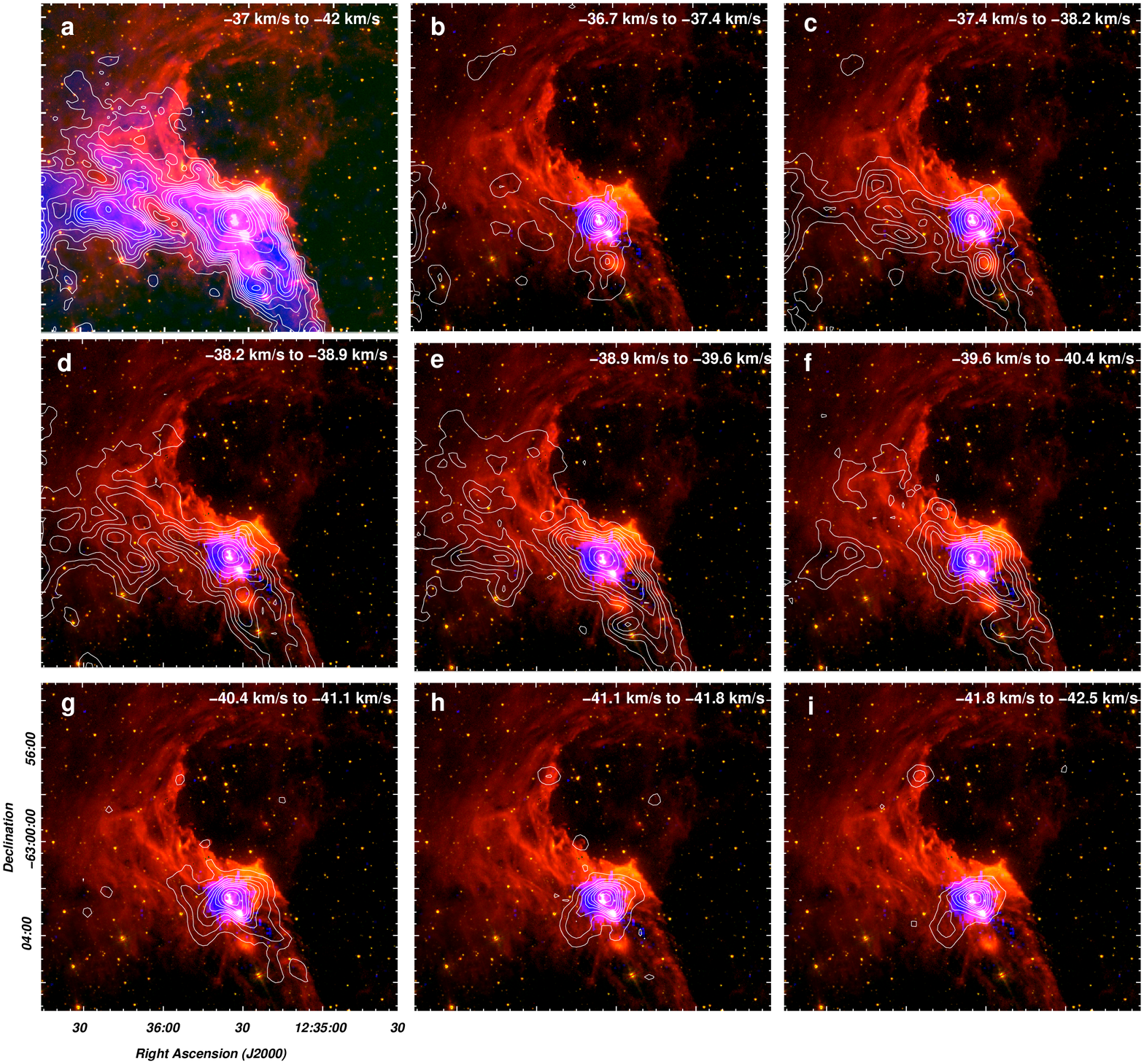}
      \caption{{\bf \it Panel a:} Emission distribution of the $^{13}$CO(2-1) line in the total velocity range from --37 to --42 \kms\ (white contours and blue color scale) superimposed on the IRAC-GLIMPSE images at 8.0 $\mu$m and 5.8 $\mu$m (red and green color scales, respectively). The contours levels are the same than those shown in the middle panel of Fig.\ref{coprom}.  {\it Panels b to i:} Channel maps of the \cob\ line emission in velocity intervals of 0.7 \kms\ (white contours) in the velocity range from --36.7 \kms to   --42.5 \kms\ superimposed on the 8.0 $\mu$m and 5.8 $\mu$m  (IRAC-GLIMPSE) emissions (red and green color scales)  and 870 $\mu$m ATLASGAL emission (blue color scale) inside the 3 rms  limit. Contours levels are 0.9 ($\sim$ 5.5 rms), 2.6, 4.4, 6.2, 7.9, 10, 12, 16, and 20  K \kms. The velocity interval is indicated in the top right corner of each panel.  }
         \label{mosaico}
   \end{figure*}

In Fig.~\ref{coprom}, we show the spatial distribution of the \coa, \cob, and \coc\ lines in the velocity interval from  $\sim$ $-$42 \kms\ to $-$37 \kms.  As expected  according to the IR emission distribution (see Fig.~\ref{irac-sc}), the bulk of  the molecular  emission appears concentrated toward the eastern and southern borders  of the IR nebula, with no significant molecular emission detected toward the western and northeastern regions. The morphology and location of component A, H$\alpha$ emission (see Fig.~\ref{irac-sc}), and radio continuum emission (see Sect. \ref{ioniz})  suggest that the ionized gas is expanding against the molecular cloud in the eastern and southern borders. The molecular emission at the southern border of the bubble seems to be tracing the densest areas, which are probably sculpted by the action of the   \hii\ region.  The morphology and location  of component A with respect to the near IR emission is similar to many other in Galactic IR bubbles found in the literature (e.g., \citealt{zav06,deh08,deh09,and15,liu15,cap16,dur17,dev18}).  For a better visualization and comparison with the IR emission, in panel {\it a} of Fig.~\ref{mosaico} we show the emission of \cob\ of component A superimposed on the IRAC-GLIMPSE emissions at 8.0 and 5.8 $\mu$m.

From Fig.~\ref{coprom}, it can also be seen that  the CO emission of component A is more extended than that of $^{13}$CO and C$^{18}$O. This indicates that  $^{13}$CO and C$^{18}$O lines, optically thinner than that of CO (see below), are actually tracing the distribution of the densest molecular gas surrounding  the bubble nebula. In order to perform a more detailed analysis of the dense molecular gas in component A,  we have roughly identified six molecular condensations in the \cob\ emission using the \coc\ emission  (which traces the denser gas) as a reference.  The identification was done using the 1.5 K \kms\ contour level in the \cob\ emission. The condensations are labeled in Fig.~\ref{coprom} as MC1, MC2, MC3, MC4, MC5,  and MC6.  These condensations  are the places where star formation is likely to be taking place (see Sect. \ref{stelyprot}).  In  panels {\it b} to {\it i} of Fig.~\ref{mosaico}, we show the channel maps of $^{13}$CO in the velocity range from  $-$36.7 \kms\  to $-$42.5 \kms\ in intervals of 0.7 \kms. In order to compare the molecular emission distribution with the warm and  cold dust emission, the $^{13}$CO  emission was superimposed on the 8.0 and 5.8 $\mu$m (IRAC-GLIMPSE) and the 870 $\mu$m (ATLASGAL)  maps. Figure~\ref{mosaico} shows that MC3, located at \radec = (12\h 35\m 35\s , $-$63\gra02\arcmin25\arcsec), is the only  molecular condensation that is present in the entire velocity range. The position and size of MC3  perfectly match with those of the bright spot seen at 870 $\mu$m.   This condensation certainly represents the carbon monoxide counterpart of IRAS~12326-6245, which reveals the physical association between S169 and the IRAS source.  In the velocity interval from $-$36.7   to $-$38.2 \kms\ condensation MC1 becomes noticeable, reaching its peak temperature in the velocity interval from $-$38.9  to $-$39.6 \kms, close to a 8.0 $\mu$m source seen at \radec = (12\h 35\m 24\s , $-$63\gra05\arcmin42\arcsec). It can be detected till a velocity of $-$40.4 \kms. For the case of condensation MC2, it can be first noticed in the velocity interval from $-$36.7  to $-$38.9 \kms, where its emission merges with MC1. Its peak emission is observed in the velocity range from $-$37.4  to $-$38.2 \kms\ and is coincident with another 8.0 $\mu$m source seen at \radec = (12\h 35\m 29\s , $-$63\gra04\arcmin25\arcsec). Condensation MC4 becomes detectable in the velocity interval from $-$36.7 to $-$37.4 \kms, reaching to its maximum peak emission in the velocity interval from $-$38.2 to $-$38.9 \kms. It can be noticed till a velocity of $-$39.6 \kms\ although its emission appears blended with MC2. Regarding condensations  MC5 and MC6, they become detectable in the velocity range from $-$36.7  to $-$37.4 \kms. Their emission distribution is quite irregular and achieve their peak temperatures in the velocity range from $-$38.2  to $-$38.9 \kms. Condensation MC4 is detectable till a velocity of $-$39.6 \kms, while  MC5 till a velocity of $-$40.4 \kms.

Figure~\ref{mosaico} also shows a strong morphological correlation between the brightest parts of the PDR (described in Sect.~\ref{intro}) and the molecular condensations MC3, MC4, and MC5, especially  at velocities between --38.2 to --38.9 \kms. In particular, it is noticeable  how  the bright emission feature detected at 5.8 and 8.0 $\mu$m borders the area of MC3, which are likely  to be exposed to stellar radiation.

To estimate the physical properties of the $^{13}$CO condensations, we obtained six $^{13}$CO emission maps (not shown here) integrated in the velocity interval in which each condensation is detected.
The deconvolved effective radius of the condensations derived from the $^{13}$CO line, $R_{\rm D}^{\rm ^{13}CO}$,  is calculated as:
\begin{equation}
  R_{\rm D}^{\rm ^{13}CO} =   \sqrt{R^2_{\rm eff} - ({\rm HPBW}/2)^2},
\end{equation} 
where  $R_{\rm eff}$  is the effective radii of the condensation ($R_{\rm eff}$ = $\sqrt{A_{\rm cond}/\pi}$), being $A_{\rm cond}$ the area of the condensation,  and HPBW is the half-power beam width of the instrument. To estimate $R_{\rm D}^{\rm ^{13}CO}$ and other parameters, we  adopted a distance of 2.03$^{+0.77}_{-0.61}$ kpc  (see Sect.~\ref{dist}). Then we obtained the effective radii of 0.40, 0.27, 0.94, 0.49, 0.75, and 0.86 pc for MC1, MC2, MC3, MC4,  MC5, and MC6, respectively.

Assuming that all rotational levels are thermalized with the same excitation temperature (LTE) and that the emission is optically thick, we derived the  excitation temperature of each molecular condensation, $T_{\rm exc}$,   from the \coa\  line  using 
\begin{equation}
T_{\rm peak} = T_{12}^*\left[\left(e^{\frac{T_{12}^*} {T_{\rm exc}}}-1\right)^{-1}
-\left(e^{\frac{T_{12}^*} {T_{\rm bg}}}-1\right)^{-1}\right]
\label{texc12co}
,\end{equation}
where $T_{\rm 12}^*$ = $h \nu_{\rm 12} / k$, being $\nu_{\rm 12}$ the frequency of the $^{12}$CO(2-1) line,  and $T_{\rm bg} = 2.7$ K. 
To obtain the peak main beam temperature of the \coa, \cob, and \coc\ lines $(T_{\rm peak}),$ we used Gaussian fits over the spectra profiles obtained integrating in the direction of  the peak emission of the $^{13}$CO condensations (in all the cases coincident with those of the C$^{18}$O emission) over an area equivalent to the area of the beam. It is worth to point out that a single Gaussian component was  considered for MC3, which shows self-absorption in the spectrum of  \coa\ at the velocity of the peak emission of \cob\ and \coc\ \citep{hen00,ded11}. Excitation temperatures derived with Eq.~\ref{texc12co} are  then 21.3 K, 33.8 K, 47.8 K, 21.2 K, and 16.0 K, and 16.8 K for MC1, MC2, MC3, MC4, MC5, and MC6,  respectively. 

\begin{table*}
\caption{Observed parameters obtained from the emission lines and physical properties derived for the molecular condensations MC1 to MC6}
\label{tabla-propiedades}
\centering
\begin{tabular}{|c|c|c|c|c|c|c|c|c|}
\hline
 &  $T_{\rm peak}$  &  $\int T_{\rm mb}\ dv$  &  $\Delta$v  &                              $\tau$   &  $N(^{13}$CO)           &  $N$(C$^{18}$O)     &  $N({\rm H_2})$ &  $M({\rm H_2})$     \\
 &        (K)       &    (K \kms)              &  (\kms)    &                                       &   (cm$^{-2}$)  &   (cm$^{-2}$)   &  (cm$^{-2}$)    &  ($M_{\odot}$) \\

       &  CO\  $^{13}$CO\  C$^{18}$O  &     $^{13}$CO\  C$^{18}$O   &  CO\  $^{13}$CO  &    CO\  $^{13}$CO\  C$^{18}$O   &           &            &          &                       \\

\hline
\hline
MC1  & 16.1\ \ \ 6.8\ \ \ 0.8      &  15.9\ \ \ 0.7  & 2.75\ 1.49    &           16.51\ \ 0.54\ \  0.05  &  2.8$\times$10$^{16}$  &  1.0$\times$10$^{15}$  &     (2.0/0.6)$\times$10$^{22}$ &  300    \\
     &                             &                 &               &           &                         &        &          &   \\
MC2  & 28.4\ \ \ 10.0\ \ \ 1.3      & 7.8\ \  \ 1.0  & 2.85\ 1.91    &         16.32\ \ 0.43\ \ 0.04   & 1.6$\times$10$^{16}$   & 1.6$\times$10$^{15}$  & (1.2/0.9)$\times$10$^{22}$   &  90    \\
     &                              &                &               &           &                         &        &          &        \\
MC3  & 42.2\ \ \ 23.3\ \ \ 7.1     &  32.8\ \ \ 8.1   & 7.51\ 5.85    &         36.2\ \ 0.83\ \ 0.19   & 9.8$\times$10$^{16}$   &  1.8$\times$10$^{16}$ & (1.0/0.7)$\times$10$^{23}$   & 5500     \\
     &                             &                  &               &           &                         &        &           &        \\
MC4  &  15.9\ \ \ 7.4\ \ \ 2.1    & 9.9\ \ \ 1.2    &  2.83\ 2.16   &         26.38\ \ 0.61\ \ 0.13   & 1.8$\times$10$^{16}$  &  1.1$\times$10$^{15}$ & (1.3/0.7)$\times$10$^{22}$   &  280   \\
&&&&&&&&\\
MC5  &  10.9\ \ \ 6.9\ \ \ 1.6    &  6.1\ \ \ 0.8   &  3.18\ 1.59   &         27.76\ \ 0.98\ \ 0.15   &  1.2$\times$10$^{16}$  &  1.1$\times$10$^{15}$ & (8.5/6.5)$\times$10$^{21}$   &  420 \\
&&&&&&&&\\
MC6  &  11.7\ \ \ 8.4\ \ \ 1.3    &  7.1\ \ \ 1.1   &  2.55\ 1.65   &            44.36\ \ 1.21\ \ 0.11  & 1.6$\times$10$^{16}$  &  1.5$\times$10$^{15}$ & (1.2/0.9)$\times$10$^{22}$  &  720 \\
\hline
\end{tabular} 
\end{table*}

The optical depths ($\tau$) of the  \cob\ and \coc\ lines, denoted in the next equations as $\tau^{13}$ and $\tau^{18}$, respectively,  were obtained  assuming that the excitation temperature is the same for \coa, \cob, and \coc,  and using the expressions

\begin{equation}
\tau^{13}=-{\rm ln}\left[1-\frac{T_{\rm peak} {\rm ^{13}CO}}{T_{13}^*}
\left[\left(e^{\frac{T_{13}^*} {T_{\rm exc}}}-1
\right)^{-1}-\left(e^{\frac{T_{13}^*} {T_{\rm bg}}}-1\right)^{-1}\right]^{-1}\right]
\label{tau13co}
\end{equation}
and
\begin{equation}
\tau^{18}=-{\rm ln}\left[1-\frac{T_{\rm peak} {\rm C^{18}O}}{T_{18}^*}
\left[\left(e^{\frac{T_{18}^*} {T_{\rm exc}}}-1
\right)^{-1}-\left(e^{\frac{T_{18}^*} {T_{\rm bg}}}-1\right)^{-1}\right]^{-1}\right]
\label{tauc18o}
,\end{equation}
\noindent where $T_{\rm 13}^*$ = $h \nu_{\rm 13} / k$, $T_{\rm 18}^*$ = $h \nu_{\rm 18} / k$,      with $\nu_{\rm 13}$  and $\nu_{\rm 18}$  the frequencies of the \cob\ and \coc\ lines, respectively. We can also estimate the optical depth of the CO(2-1) line  from the $^{13}$CO(2-1) line using  
\begin{equation}
\quad \tau^{12} =\  \left[\frac{\nu^{13}}{\nu^{12}}  \right]^2\ \times  \left[\frac{\Delta {\rm v}^{13}} {\Delta   {\rm v}^{12}} \right]\ \times  \left[\frac{\rm CO}{^{13}{\rm CO}} \right]\   \tau^{13}
\label{tau12}
,\end{equation}
where   CO/$^{13}$CO is the isotope ratio (assumed to be $\sim$ 62; \citealt{lp93});  $\Delta{\rm v}^{13}$ and $\Delta{\rm v}^{12}$ are defined as the full width half  maximum (FWHM) of the spectra  of the $^{13}$CO and CO lines, respectively, which are derived by using a single Gaussian fitting  (FWHM = 2 $\times$ $\sqrt{2\ ln2}$ $\times$ $\sigma_{\rm gauss}$).

In LTE,  the $^{13}$CO  and C$^{18}$O column densities   can be estimated from the \cob\ and \coc\  line using
\begin{equation}
N({\rm^{13}CO})=3.23\times 10^{14} \left[\frac{e^{  \frac{T_{13}^*}{T_{\rm exc}}  }}{1 -e^{-  \frac{T_{13}^*}{T_{\rm exc}}   } }\right]  T_{\rm exc} \int \tau^{13}\   dv \ \  \textrm{ (cm$^{-2}$)} 
\label{n13co}
\end{equation}
and
\begin{equation}
N({\rm C^{18}O})=3.21\times 10^{14} \left[\frac{e^{  \frac{T_{18}^*}{T_{\rm exc}}  }}{1 -e^{-  \frac{T_{18}^*}{T_{\rm exc}}   } }\right]  T_{\rm exc} \int \tau^{18}\   dv \ \  \textrm{ (cm$^{-2}$)}
\label{nc18o}
.\end{equation}
\noindent The integral in Eqs.~\ref{n13co}  and  \ref{nc18o}  can be approximated by
\begin{equation}
\ \ T_{\rm exc} \int{\tau\  dv \approx\ \frac{\tau}{1-e^{(-\tau)}}
\int{T_{\rm mb}}}\ \ d{\rm v}.
\label{integral}
\end{equation}
This approximation helps to eliminate to some extent optical depth effects and is good within 15\% for $\tau <$  2 (\citealt{rw04}). Bearing in mind the  values of $\tau^{\rm 13}$ obtained,  the approximation is appropriate for our region.  Estimations of the H$_{\rm 2}$ column density, $N$(H$_{\rm 2}$), were obtained from both the $^{13}{\rm CO}$ and  C$^{18}$O column densities, and  adopting  abundances  $X$[$^{13}$CO] = 7.1$\times$10$^{5}$ and $X$[C$^{18}$O] = 5.9$\times$10$^{6}$ \citep{frer82}.  For the optical depth and column density, we estimated uncertainties of about 20 $\%$ and 25 $\%$, respectively, arising mostly from calibration uncertainties of about 20 $\%$ \citep{dum10}.

The total hydrogen mass  of each clump  was calculated using
\begin{equation}\label{eq:masa}
\ \ {\rm M(H_2)}\ =\ (m_{\rm sun})^{-1}\ \mu\ m_{\rm H}\  \Omega \ {\it N}(\rm H_2)\ {\it d}^2 \quad \quad \quad \textrm{(M$_{\odot}$)}
,\end{equation}
where $m_{\rm sun}$ is the solar mass ($\sim$2$\times$10$^{33}$ g), $\mu$ is the mean molecular weight, which is assumed to be equal to 2.76 after allowing for a relative helium abundance of 25\% by mass \citep{yama99}, $m_{\rm H}$ is the hydrogen atom mass ($\sim$1.67$\times$10$^{-24}$ g),  $\Omega$ is the solid angle of the $^{13}$CO emission  expressed in sr, $d$ is the adopted distance expressed in cm, and $N$(H$_{\rm 2}$) is the H$_{\rm 2}$ column density   obtained from  $N({\rm^{13}CO})$.    
Uncertainties in molecular masses are about 60 $\%$, and originate mainly from a distance uncertainty of about 30 $\%$ (see Sect.~\ref{dist}). The derived physical parameters are presented in Table~\ref{tabla-propiedades}. 

\begin{figure}
 \centering    
   \includegraphics[width=7.5cm]{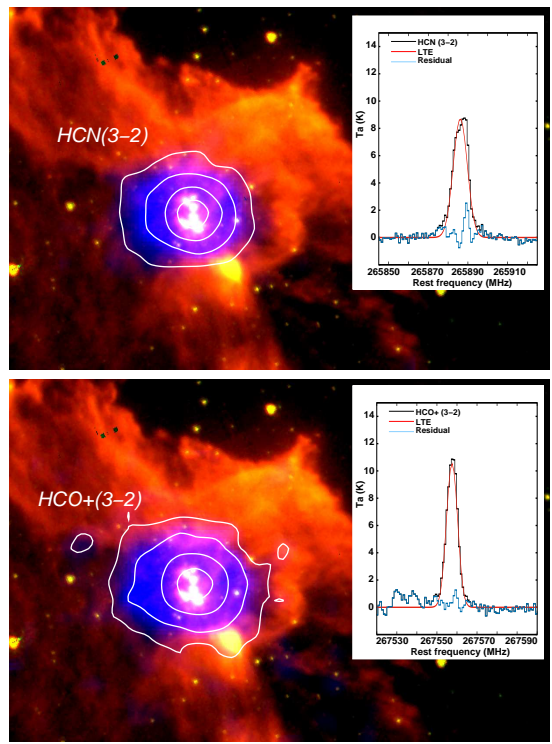}
    \caption{{\it Upper panel:} Spectral line and LTE model of the \hcn\ line obtained from LTE-MCMC calculations assuming $T_{\rm exc}$=150~K (see text), with a resulting density of $N$=(1.9 $\pm$ 0.5) $\times$ 10$^{14}$~cm$^{-2}$. The blue line indicates the residual obtained from the subtraction between the observation and LTE model. In colors: Spatial distribution of the \hcn\ emission in the velocity interval from $-$47.6   to $-$29.2 \kms (color blue and white contours) superimposed on  the IRAC  8  and 5.8 $\mu$m emissions (red and green colors).   {\it Lower panel:} Same as above for the \hco\ line, assuming $T_{\rm exc}$=50~K, with a  resulting density of $N$=(5.2 $\pm$ 0.1) $\times$ 10$^{13}$~cm$^{-2}$. The  \hco\ emission distribution is in the velocity interval from $-$44.8  to  $-$32.5 \kms. The contour levels for the \hcn\  are 0.25 ($\sim$5 rms), 0.85, 2.2, and  4.0 \hbox{K \kms}; and for the \hco\ line, they are 0.28  ($\sim$ 5 rms), 0.8, 2.4, and 5  \hbox{K \kms}   }
\label{hcoyhcn}
\end{figure}

In order to perform a study of the molecular gas in IRAS~12326-6245, we also present an analysis for MC3 using the high-density tracer lines \hcn\ and \hco. In Fig.~\ref{hcoyhcn}, we show the  emission  of these lines  in the velocity intervals from $-$47.6 to $-$29.2 \kms\ and $-$44.8 to $-$32.5 \kms, respectively. Their emissions show a good morphological correspondence with that at 870 $\mu$m (see Fig.~\ref{mosaico}) and  seem to be delineating  an obscured region in the 8.0 and 5.8 $\mu$m emissions.  They are certainly the HCN and HCO$^+$ counterparts of the molecular condensation MC3. In order to estimate the  HCN and HCO$^+$   column densities, we used LTE models numerically approximated with the Markov chain Monte Carlo (MCMC) method \citep{Foreman2013}. The lines in LTE were simulated and compared with the observations  (e.g., \citealt{Comito2005,Moller2017}). The solutions presented were obtained from the averaged spectra considering different offsets since the individual solutions for each one of them were similar. That is a consequence of the homogeneous emission traced by the species analyzed here.  In Fig.~\ref{hcoyhcn}, we show the average spectra (in antenna temperature)  of \hcn\ and \hco.  The results presented here were obtained by taking into account those reported in IRAS\,12326-6245 by \citet{ded11}, who assumed temperature values of 150~K and 50~K for the gas conditions of species in an apparent hot core region and extended envelope, respectively.  For the sake of the analysis, we also included the optically thin lines of \cob\ and \coc. Then, assuming an excitation temperature of $T_{\rm exc}$=50 K for the \hco\ line, we obtained a column density of  $N$(HCO$^+$) = (5.2 $\pm$ 0.1) $\times$ 10$^{13}$~cm$^{-2}$. For the \cob\ and \coc\ lines, also adopting $T_{\rm exc}$=50 K, we  obtained $N$($^{13}$CO) = (2.01 $\pm$ 0.04) $\times$ 10$^{17}$~cm$^{-2}$ and  $N$(C$^{18}$O) = (3.9 $\pm$ 0.1) $\times$ 10$^{16}$~cm$^{-2}$.  For the case of the \hcn\ line, LTE models suggest a physical condition related with the hot core region. Based on the hypothesis stating that $T_{\rm exc}$=150~K, we derived $N$(HCN) = (1.9 $\pm$ 0.5) $\times$ 10$^{14}$~cm$^{-2}$. Concerning the uncertainties, the results presented here are approximations that should be considered (at least) within the calibration uncertainty since several assumptions were adopted for single transitions. In order to inspect the quality of the fits, in Fig.~\ref{hcoyhcn} we also exhibit the residuals after subtracting the emission lines  from their LTE models.

\subsection{Components B and C}\label{compbyc}

As mentioned in Sect.~3.1, two weaker broad velocity molecular components peaking approximately at $-$25 \kms\ (component B) and $-$17 \kms\ (component C) were also detected in S169. These components are hardly detected in the $^{13}$CO emission and not detected at all in the C$^{18}$O emission, which  certainly indicates that they are mostly composed by low-density  gas. 
\begin{figure}
   \centering
   \includegraphics[width=8cm]{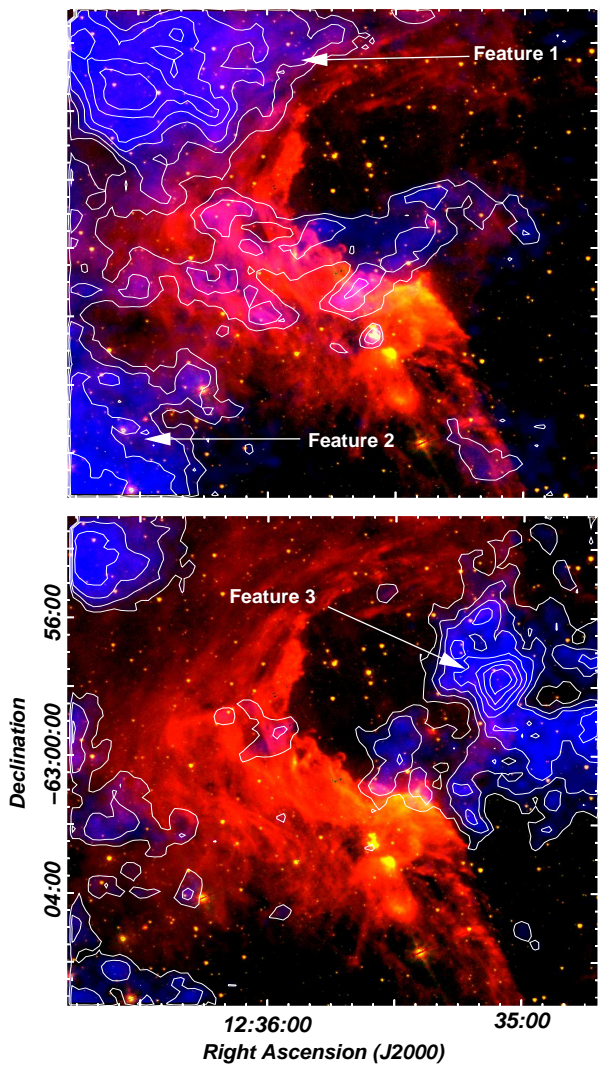}
       \caption{ {\it Upper panel:} Spatial distribution of the CO  emission (blue color scale and white contours) in the velocity interval from $-$28.9 \kms\ to $-$18.9 \kms\ (component B) superimposed on the IRAC  8 $\mu$m and 5.8 $\mu$m emissions (red and green color scales).  The contour levels start from  0.95 \hbox{K \kms} ($\sim$ 9 rms) in steps of 0.5 \hbox{K \kms}. {\it Lower panel:} Spatial distribution of the CO  emission  in the  velocity interval from $-$18.2 \kms\ to $-$12.7 \kms\ (component C). The contour levels start from  0.85 \hbox{K \kms} ($\sim$ 7 rms) in steps of 0.4 \hbox{K \kms}. }  
    \label{combyc}
   \end{figure}
In order to analyze their spatial distribution, we constructed two CO emission maps integrated in the velocity intervals from $-$28.9 to $-$18.9 \kms\ (component B) and from $-$18.2 to $-$12.7 \kms\ (component C). They are shown in Fig.~\ref{combyc} superimposed onto the 8.0 and 5.8 $\mu$m emissions. For the  case of component B, three molecular features are seen projected towards different regions of the nebula. The brightest feature (hereafter, feature 1) is seen approximately centered at \radec = (12\h 36\m 34\s, $-$62\gra55\arcmin13\arcsec) perfectly delineating the northeastern and northern borders of the IR nebula. As can be seen from the figure,  the brightest regions of the molecular gas appear projected onto the faintest regions of the IR emission, which suggests that a PDR was formed over the surface of feature 1 (viewed from the side).   Another feature is seen centered approximately at \radec = (12\h 36\m 40\s , $-$63\gra06\arcmin00\arcsec) (feature 2). This feature appears bordering the southeastern edge of the nebula, also projected onto the faintest regions of the IR nebula. The third feature is seen projected along the center of the nebula approximately at Dec=$-$62\gra58\arcmin00\arcsec\ and is still noticeable at more negative velocities. Since this feature shows no morphological correspondence with any region  of the IR nebula, a physical link with S169 cannot be suggested. Probably, it is an unconnected molecular structure located behind the nebula since no absorption features can be noticed along the  brightest regions of the IR emission. Regarding component C, only one molecular structure (feature 3) stands out, peaking approximately at \radec = (12\h 35\m 05\s , $-$62\gra58\arcmin00\arcsec). This feature shows a good morphological correspondence with the faint IR  arclike filament seen at the western section of the nebula at RA =  12\h 35\m 10\s.  This correspondence is also well-displayed in Fig.~\ref{ysos-exc}.  Another structure has been noted at the northeastern region of the nebula, however\ it represents the remains of feature 1 at slightly more positive velocities.
\begin{table}
\caption{Physical properties estimated for the CO feature 1, feature 2, and  feature 3}
\centering
\begin{tabular}{|c|c|c|c|c|}
\hline
         & $\Omega$     &   $\int{T_{\rm  mb}({\rm CO})} \ d{\rm v}$     &   $N({\rm H_2})$  & $M(\rm H_2)$    \\
Feature & (10$^{-6}$ sr)       &  (K \kms)          &  (10$^{20}$cm$^{-2}$)    &  ($M_{\odot}$)  \\                
\hline 
\hline
  1      &    2.4     &  14.9       &  2.4     &  52  \\
&&&&\\ 
 2       &    1.3     &    12.3     &   2.0      &  24   \\
&&&&\\
3        &    2.5     &   7.9      &  1.3    &   29\\
\hline
\end{tabular}
\label{prop-features} 
\end{table}

Since no discernible emission of the \cob\  line  is detected for features 1, 2, and 3,  in order to estimate their  column density and mass,   we use the relation between the H$_2$ integrated column density  and the CO integrated emission:
\begin{equation} 
  N({\rm H_2})\ =\  X\ \times \int{T_{\rm  mb}({\rm CO})} \ d{\rm v}, 
\label{NH2}
\end{equation}  
 where $X$ is an empirical factor that has been shown to be roughly constant for the \hbox{$^{12}$CO(1-0)} line in  Galactic molecular clouds and  lies in the range of \hbox{(1 - 3)} $\times$ 10$^{20}$ cm$^{-2}$ (K \kms)$^{-1}$, as estimated by the virial theorem and $\gamma$-ray emission   \citep{blo86,sol87,dig96,sm96}. In this paper, we adopt $X$ = 1.6 $\times$ 10$^{20}$ cm$^{-2}$ (K \kms)$^{-1}$ \citep{hun97}. The integrated emission $\int{T_{\rm  mb}({\rm CO})} \ d{\rm v}$ was calculated adopting an area determined by the first contour level indicated in Fig.~\ref{combyc}. For the column density, we estimated an uncertainty of about 50 $\%$, arising mostly from the factor $X$ and calibration uncertainties. The mass of the features were then estimated using Eq.~\ref{eq:masa}. The obtained parameters are shown in Table~\ref{prop-features}. It is worthwhile to point out that the area and mass obtained for  features 1 and 2  are likely lower limits since these features seem to be part of a larger structure (extending beyond  the area covered in our APEX observations), which probably corresponds to the parental molecular cloud in which the IR bubble was formed.

\subsection{Submillimeter dust emission}\label{submill}

\begin{figure}   
   \centering
   \includegraphics[width=8.2cm]{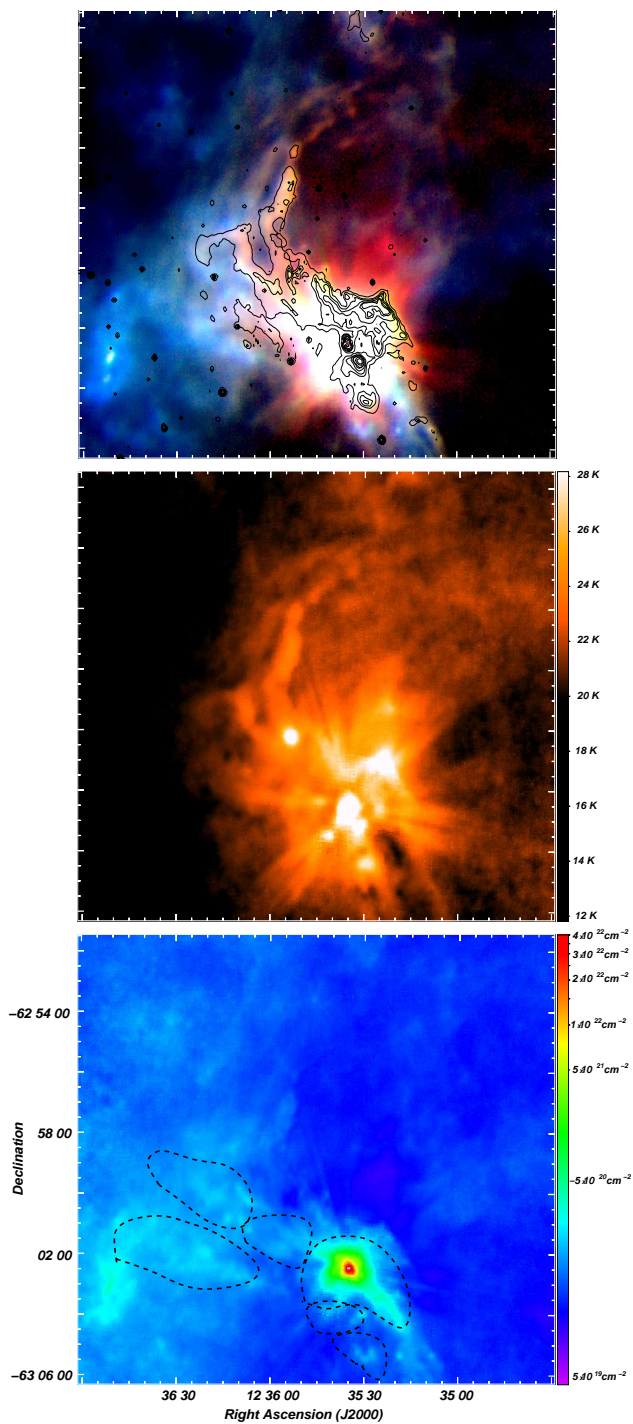}
      \caption{{\it Upper panel:} Three-color composite image of S169: {\it Herschel} PACS  70 and 160 $\mu$m emission are  in red and green, respectively. {\it Herschel} SPIRE 350 $\mu$m emission is in blue. Black contours underline de IRAC-GLIMPSE 8 $\mu$m emission at 35, 55, 75, 105, 150, 170, 300, and 500 MJy sr$^{-1}$. {\it Middle panel:} Dust  temperature map  derived from the 70 and 160 $\mu$m emission. The color temperature scale is on the right. {\it Lower panel:} Column density map obtained from the 160 $\mu$m emission.  Dashed black contours indicate the location of molecular concentrations MC1 to MC6 for orientation. } 
\label{spire-pacs}
\end{figure}

In the upper panel of Fig.~\ref{spire-pacs}, we present a three-color composite image of S169  as seen by {\it Herschel} images. The 70 and 160 $\mu$m emissions (red and green color scales, respectively) underline the emission of the warmer dust close to the ionized gas. As can be seen from the  figure, the emission at 70 $\mu$m has a good morphological correspondence with the IRAC 8 $\mu$m emission (black contours), although the latter exhibits a sharper edge towards the center of the nebula (see also Fig.\ref{irac-sc}), which is likely due to the destruction of the PAH molecules by the ionization front. The 70 $\mu$m emission, in contrast, is  more diffuse (partly because of the lower angular resolution) and extends towards the center of the nebula, which indicates that it is also tracing the warm dust that still remains mixed  with the ionized gas. The same feature can be observed in the IRAC-GLIMPSE 24 $\mu$m emission (not shown here). The 350 $\mu$m  emission (blue color scale) seems to be underlining the emission from the  dust in the outer parts of the nebula. It is very likely that the emission distribution at this wavelength is mostly tracing the distribution of the cold dust immersed in the parental molecular cloud placed at the eastern border of the nebula, which is  partially disclosed by  features 1 and 2  (see Fig.~\ref{combyc}).

To study the distribution of the dust  temperature in the nebula, we constructed  a  temperature map from the ratio of the observed fluxes in two {\it Herschel} bands. Since a high  level of radiative feedback from powering massive stars is usually observed in \hii\ regions/IR bubbles, it is reasonable to assume that the molecular gas and dust are  not too cold. Then we used  the ratio of the 70 $\mu$m versus 160 $\mu$m maps, which are better suited to measure color temperatures up to $\sim$ 80 K. Furthermore, the  angular resolution of the temperature map obtained using these bands  is the highest resolution achievable  from the {\it Herschel} maps. The 70\,$\mu$m map was smoothed down to the angular resolution  of the 160\,$\mu$m data. Then the  temperature map was constructed as the inverse function of the ratio  70 to 160\,$\mu$m maps  that is, $T_{\rm c}=f^{-1}_{(T)}$. Assuming a dust emissivity following a power law of \hbox{$\kappa_{\nu}$ $\propto$ $\nu^{\beta}$},  with ${\beta}$ as the spectral index of the thermal dust emission (assumed to be $\beta$ = 2),   in the optically thin thermal dust emission regime, $f_{(T)}$ takes the parametric form: 
\begin{equation}
\qquad f_{(T)} = \frac{S_{70}}{S_{160}} = \frac{B(70,T)}{B(160,T)} \left( \frac{70}{160} \right) ^{\beta}
,\end{equation}
 where $B(70,T)$ and $B(160,T)$   make up the blackbody Planck function for a temperature $T$ at wavelengths 70  and 160 $\mu$m, respectively. The assumption that the emission is optically thin  is further justified from the derived column densities in the nebula  (see \citealt{pre12} for details).      From the obtained map (middle panel of Fig.~\ref{spire-pacs}),  dust temperatures between $\sim$ 20  and 30 K can be discerned. These temperatures are commonly observed in the closest regions of IR bubble nebulae (e.g., \citealt{and12,dur15,fig17}).  Warmer dust is seen in the region of  the denser molecular condensations (MC1, MC2, MC3, and MC4) and the border of the PDR, particularly on two spots at  \radec = (12\h 35\m 24\s , $-$63\gra00\arcmin56\arcsec) and \radec = (12\h 35\m 35\s , $-$63\gra02\arcmin29\arcsec)    where the temperature achieves $\sim$ 30 K. The first spot is coincident with a  bright rim placed over the southern region of the PDR   and a peak in the radio continuum emission (see Fig.\ref{843} in Sect.~\ref{ioniz}), which suggests external heating of the dust. The second spot is coincident with the molecular condensation MC3 (IRAS12326-6245) which is very likely  to indicate that the dust is also internally heated. Not surprisingly, lower temperatures (12 - 18 K) are seen in the region of feature 3 and behind the PDR, coincident with the position of features 1 and 2.

 With the temperature map, we converted the surface brightness map at 160 $\mu$m  into a beam averaged column density map using \citep{hil83}:
\begin{equation}
\qquad  N({\rm H_2})\   =\ R\ \ \frac{S_{\rm 160}}{\Omega_{\rm beam}\  \kappa_{160}\ \mu\ m_{\rm H}\  B(160, T_{\rm dust})}
\label{cmd}
,\end{equation} 
where $\Omega_{\rm beam}$ is the beam solid angle \hbox{($\pi$ $\theta_{\rm HPBW}^2$ / 4\ ln~2)}, $R$ is the gas-to-dust ratio (assumed to be 100), and $\kappa_{160}$ is the dust opacity per unit mass at 160 $\mu$m assumed to be 40.5 cm$^2$ g$^{-1}$  \citep{oss94}. The resulting map (presented in the lower panel of Fig.~\ref{spire-pacs}) gives an overall view of the density distribution in the whole region of the nebula. Clearly, denser regions are observed behind the PDR in the region of component A, especially  towards IRAS12326-6245, where the column density achieves peak values up to $\sim$ 4$\times$10$^{22}$ cm$^{-2}$. 

 As shown in Figs.~\ref{irac-sc} and \ref{mosaico}, the source IRAS\,12326-6245    is very bright in the continuum emission at 870 $\mu$m (no emission at this wavelength is detected above the 3~rms limit outside the IRAS source). The emission at 870 $\mu$m  is usually optically thin and dominated by the thermal emission from dust contained in dense material (e.g.,  dense molecular cores or filaments). Then we used the image of ATLASGAL  to more confidently derive  the beam averaged density in IRAS\,12326-6245 using Eq.~\ref{cmd}. For the calculations, we adopted $\kappa_{870}$ = 1.85 cm$^2$ g$^{-1}$, a value that is representative for relatively dense molecular clouds \citep{oss94,hen95} and $T_{\rm dust}$ = 30 K  and we obtained a column density of $N({\rm H_2)}$=8.2 $\times$ 10$^{23}$ cm$^{-2}$. This value is almost two orders of magnitude higher than that observed in the column density map obtained with the emission at 160 $\mu$m, which might indicate that the emission is optically thick at that wavelength. Furthermore, the temperature used to derive $N({\rm H_2)}$ in IRAS\,12326-6245 (30 K)  might have been underestimated since, as mentioned above, its emission at 160 $\mu$m might not be optically thin at such high density. This could led to an overestimation of the column density in the IRAS source.  The column density obtained from the 870 $\mu$m emission is also somewhat  higher than those derived for MC3 in Sect.~\ref{compa} from the \cob\ and \coc\ emissions.    We do keep in mind, however, that the column density values derived from continuum IR and carbon monoxide emissions are strongly dependent on the adopted abundances (dust and carbon monoxide-to H$_2$).

\section{Ionized gas}\label{ioniz}

\begin{figure}
   \centering
   \includegraphics[width=9cm]{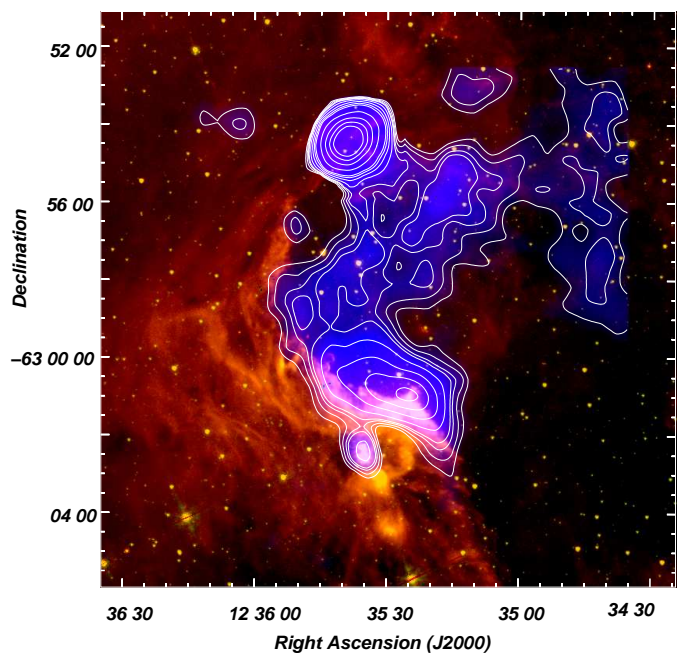}
      \caption{ Radio continuum emission distribution at 843 MHz (blue color and white contours), superimposed  on the 8 and 5.8 $\mu$m emission (red and green color). Contours levels are 3.1 ($\sim$ 3 rms), 4.5, 6.3, 7.6, 12, 20, 30, 40, 100, 200, 300, and 400 \mjyb.     }
         \label{843}
   \end{figure}

 In Fig.~\ref{843}, we show the radio continuum emission map at 843 MHz overlaid on the 8.0 and 5.8 $\mu$m emissions. Three bright structures stand out in the direction of the nebula. The brightest one (S$_{843}$ = 985  mJy), which does not have a discernible counterpart in the H$\alpha$ emission,  is observed towards the norther border of the nebula at \radec=(12\h 35\m 39\s , $-$62\gra54\arcmin26\arcsec). Using a radio continuum image from the SGPS (not shown here), we found that this compact source has an integrated flux density of S$_{1384} = 0.63$ mJy. Since the spectral index turns out to be $\alpha = -0.9$,   we conclude that this source is extragalactic and will not be taken into further consideration.

 An arc-shaped structure can be discerned towards the center of the nebula, which may be described by two main components: (i) a southern bright component with its outer edge closely delineating the PDR depicted by the 8.0 $\mu$m emission; and (ii) a weaker tail about 5$^\prime$-long extending from the northern end of the previous component towards the northwest. A number of artifacts in the radio continuum emission are obviously produced by sidelobes and grating rings from the bright extragalactic source, such as a void at \radec$\approx$(12\h 35\m 45\s , $-$62\gra56\arcmin33\arcsec ) and two radial strips connecting the compact source and the weak tail. Hence, we believe that the morphology of the arc-shaped structure is probably distorted in the region closer to the extragalactic source. However, the morphology of the unaffected part (southernmost component) suggests that the ionized gas traced by the radio continuum emission is likely to be expanding against component A and features 1 and 2, supporting the same conclusion derived in Sect.~\ref{intro} based on the H$\alpha$ emission.
 
 Using the integrated radio continuum flux  obtained for the whole arc-shaped structure  (i.e., the southern bright component and weaker tail), S$_{843}$ = 308 mJy,  we obtained its electron density and ionized mass using \citep{pan78}: 
\begin{equation}
n_{\rm e}=3.113\times10^2 \ S_{843}^{0.5}\  T_{\rm e}^{0.25}\   d^{-0.5}\  b(\nu,T_{\rm e})^{-0.5}\ \theta_R^{-1.5}\ \  {\rm cm}^{-3} 
\label{de}
\end{equation}
and
\begin{equation}
M_{\rm ion}=0.7934 \ S^{0.5}_{843}\ T^{0.25}_{\rm e}\  d^{2.5}\  b(\nu,T_{\rm e})^{-0.5}\ \theta^{1.5}_R\ \ (1+Y)^{-1}\ \  {\rm M_{\odot}},
\label{mion}
\end{equation}
where $T_{\rm e}$ is the electron temperature (assumed to be 1 $\times$ 10$^4$ K)   in units of 10$^4$ K, $\theta_{\rm R}$ is the angular radius of the source in arc minutes (assumed to be 4\arcmin), $b(\nu,T_{\rm e})$ = 1 $+$ 0.3195$\times$log($T_{\rm e}$/10$^4$ K) - 0.213$\times$log($\nu$/1 GHz), $d$ is in kpc, and $Y$ is the abundance ratio by number of He$^+$ to H$^+$. Then we obtained $n_{\rm e}$ = 15 \cm3 and $M_{\rm ion}$ = 17 \msun. 
 It is worthwhile pointing out that the presence of the extragalactic source projected on  S169  prevents the estimation of the whole radio continuum emission related to the structure.

 The number of ionizing Lyman continuum photons needed to sustain the current level of ionization in the arc-shaped structure can be calculated using \citep{chai76}:
\begin{equation}
   N_{\rm Lyc} = 0.76 \times\ 10^{47}\ T^{-0.45}_{\rm e}\  S_{843}\ \nu^{0.1}  d^2\ \ {\rm s}^{-1}
\label{nlym}
,\end{equation}
 where $T_e$ is in units of 10$^4$ K, $S_{843}$ is in Jy, $\nu$ is in GHz, and $d$ is in kpc. Then, in considering distance uncertainties, we obtained for this source a range of $N_{\rm Lyc}$ = 0.5 - 2.0   $\times$ 10$^{47}$ s$^{-1}$. We keep in mind that these values are likely   lower limits  since about 25 - 50 $\%$ of the UV photons are absorbed by interstellar dust in the H{\sc ii} region  \citep{i01}.  Furthermore, \citet{wat08}, on the basis of studies of several bubbles, concluded that  $N_{\rm Lyc}$ estimated with this method is lower than expected by about a factor of 2.   Then for the arc-shaped structure, we estimate a range of $N_{\rm Lyc}$ $\approx$  (1 - 4) $\times$ 10$^{47}$ s$^{-1}$. Adopting ionizing photon rates extracted from \citet{mar05}, we  estimate the spectral type of the ionizing star of S169 to be at most O9.5V. Alternatively, a handful of later B-type stars could also be responsible for powering the \hii\ region. The third structure detected in the 843 MHz emission is  a small bright  spot discernible in the southern region of the nebula at \radec = (12\h 35\m 35\s , $-$63\gra02\arcmin29\arcsec), which is certainly the radio continuum counterpart of IRAS\,12326-6245. Using the flux density obtained for this source  ($S_{843}$=12 mJy),  we obtained $n_{\rm e}$ = 70 \cm3, $M_{\rm ion}$ = 0.3 \msun, and $N_{\rm Lyc}$= (0.4 - 1.4) $\times$ 10$^{46}$ s$^{-1}$.  It is worthwhile pointing out that for the previous estimations, we assumed optically thin emission at 843 MHz; since that assumption cannot be verified,  estimations obtained with the Eqs.~\ref{de}, \ref{mion}, and \ref{nlym}   should be taken as the lower limits. In fact,    for the case of the radio continuum emission associated with IRAS\,12326-6245,  \citet{ded11} used cm observations obtained by  \citet{urq07} to
derive a Lyman continuum flux of 8.2$\times$10$^{48}$ s$^{-1}$, which is almost three orders of magnitude larger than the value obtained with Eq.~\ref{nlym} for the source.

In a broad sense, the distribution  of the ionized gas inside the cavity suggests an electron density gradient, with the densest material near the border of the IR nebula. This configuration suggests that the nebula is  bounded by density to the west, and  bounded by ionization to the east, north, and south. Furthermore, the structure detected to the west of the nebula along RA $\sim$ 12\h 34\m 44\s , which appears connected with the  eastern arc-shaped structure within  the 3 rms emission limit, might be indicative  of a "champagne-flow" scenario \citep{tt79}.

\section{Stellar and protostellar content in the nebula}
\label{stelyprot}
\subsection{Ionizing star candidates}
\label{exc}
With the aim of identifying possible ionizing stars in the nebula, we first inspected IR catalogs in a region of $\sim$ 7$^\prime$ in radius centered on \radec $\approx$ (12\h 35\m 28\s , $-$62\gra58\arcmin04\arcsec). Making use of GLIMPSE data, we identified Class III candidates, which are usually referred to as main-sequence (MS) young stars using color-color diagrams \citep{all04}. Then we cross-correlated these sources with sources selected from the 2MASS catalog with photometric uncertainty, $K_s <$ 0.05 mag \citep{duc01}, and we obtained $\sim$ 600 Class III sources with reliable $JHK$ photometry. Since  none of these sources were found to have spectral classification, we used the parameter $Q_{NIR}$ = $(J-H) - 1.7(H-K_s)$  following \citet{bor11} to  separate sources between early (O-B) MS stars \hbox{(-0.1$<Q_{NIR} <$0.15)} and stars that reveal IR excess \hbox{($Q_{NIR}<$ --0.1)} and are likely preMS stars or YSOs, or both. Then we obtained 60 MS candidates.

\begin{figure}
   \centering 
   \includegraphics[width=9cm]{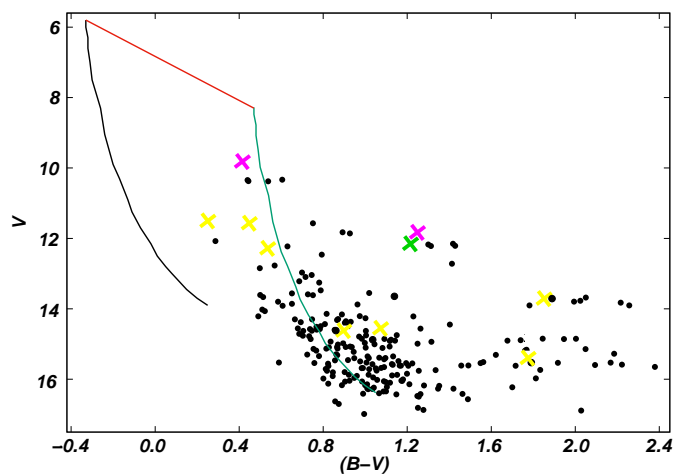}
      \caption{Color magnitude diagram of stars searched in a region centered in S169. Black dots are young spectral type stars with different reddening values (see text in Sect.~\ref{stelyprot}). Magenta, green, and yellow crosses represent the O9V, B1V, and B2V (or later) stars, all of them ionizing candidate stars displayed with the same colors in Fig. \ref{ysos-exc}. Green and black curves represent the MS shifted according to the distance modulus adopted with and without reddening, respectively. Red line indicates the normal reddening path (R$_v$ = 3.1)  }
         \label{CMagnitud} 
   \end{figure}

Next, we inspected visual catalogs in the same region. A search in the APASS and ASCC-2.5 V3 catalogs yielded 228 and 10 sources, respectively, both with V and B magnitudes. We also carried out a search in the astrometric catalog GAIA DR2 of more than 6000 sources with parallax data in the region of S169.  Thereafter, we cross-correlated all sources found with the APASS, GAIA DR2, and ASCC-2.5 catalogs, obtaining only 90 sources in common between the first two catalogs, which includes the 10 sources of the ASCC-2.5 catalog. We then  estimated the distance of visual sources employing the color-magnitude ($B-V$ vs $V$) diagram (see Fig.~\ref{CMagnitud}),  calibrated with the optical ($BV$) \citep{lan82} to adjust the theoretical MS by a distance of $\sim$ 2 kpc (see Sect.~\ref{dist}) and color excess E$(B-V)$ = 0.8 magnitudes. Thus, we were able to identify  the stars with photometric distance close to the nebula. Among these sources, only 19 had distances between 1.7 kpc and 2.3 kpc that were estimated from the parallax catalog from GAIA.  

 In the next step, we cross-correlated the 60 MS candidates obtained from the IR data with the sources previously studied with a color magnitude diagram and we found 47 sources in common. In order to better determine their spectral types, we inspected their loci in a $(J-H)$ vs. $(H-K)$ diagram adopting the MS calibration given by \citet{lan82}, \citet{cou78},  and \citet{koo83}, the absorption ratios ($r_x$ = $A_x/A_v$) given by \citet{rie85}, van der Hulst curve 15, and the color transformations from 2MASS to Koorneef data of \citet{car01}.

Finally, we kept only those stars that are projected onto the cavity of the bubble ($\sim$ 3$^\prime$ in radius centered on \radec$\approx$(12\h 35\m 28\s , $-$62\gra58\arcmin04\arcsec)). This yielded only 10 sources with spectral types  O9 (2 sources), B1 (1 source) and B2 or later (7 sources), all of them probable MS stars; it is worthwhile pointing out that only 5 of these sources belong to the group of 19 sources with distances between 1.7 kpc and 2.3 kpc obtained from GAIA.  The positions of these sources in the MC diagram and in the cavity are shown in Fig.~\ref{CMagnitud} and Fig.~\ref{ysos-exc}, respectively.

\subsection{Identification of candidate YSOs}\label{ysos}

Making use of the MSX, WISE and {\it Spitzer} point source catalogs, we look for the primary tracers of stellar formation activity  onto the molecular clouds related to S169.
\begin{figure*}[h] 
   \centering
   \includegraphics[width=16cm]{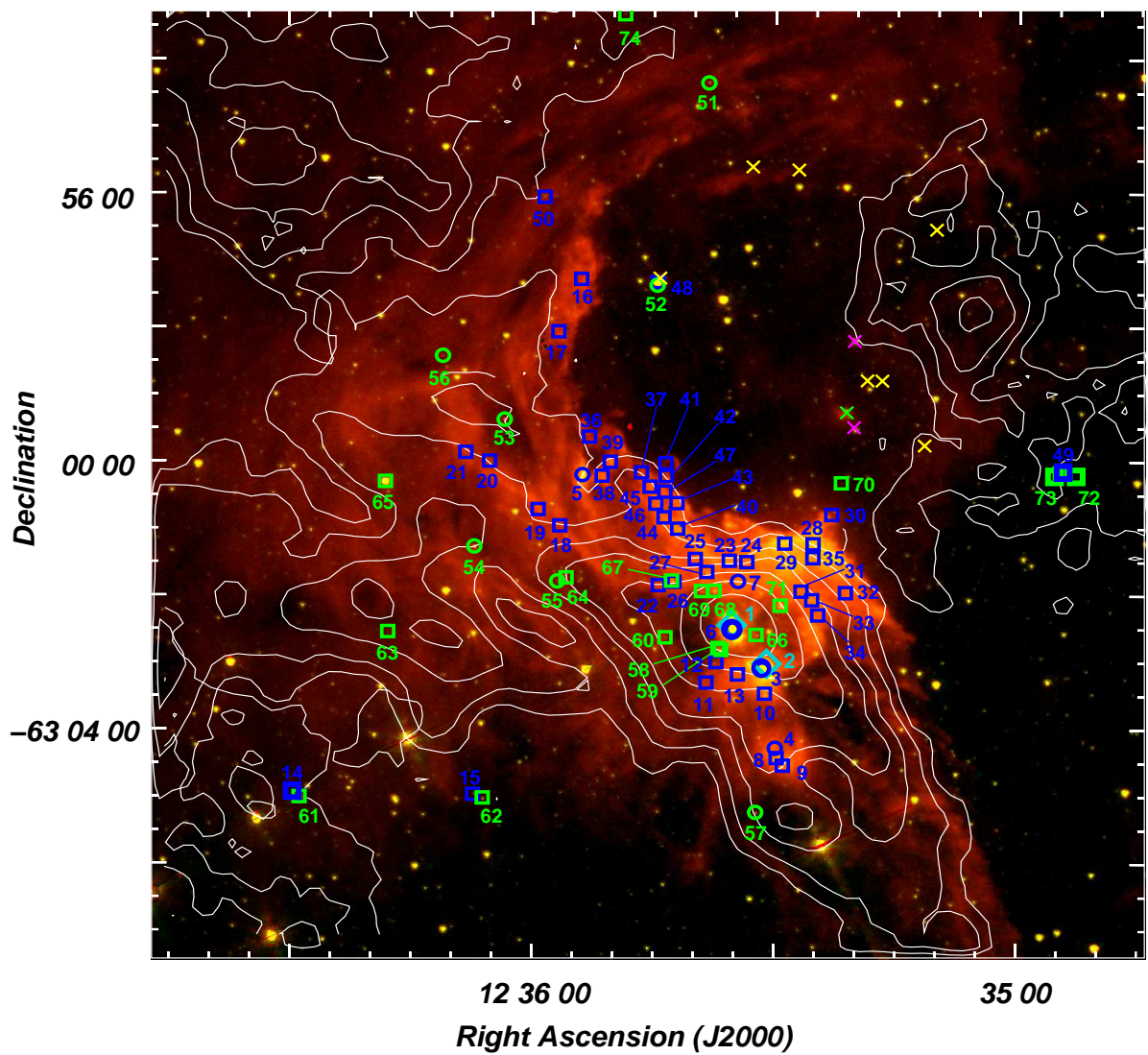}
      \caption{ MSX, WISE, and {\it Spitzer} candidate YSOs projected onto the IRAC-GLIMPSE image at 8.0  and 5.8 $\mu$m. Cyan diamonds indicate the MSX sources, blue circles and blue squares indicate WISE Class I and Class II sources, respectively, and green circles and green squares indicate the {\it Spitzer} Class I and Class II sources, respectively. The numerical references of YSOs are based on those from  Table~\ref{tysos}. The size of the symbols do not  necessary match the angular size of the sources. Ionizing candidate stars identified in Sect~\ref{exc} are also indicated by crosses in magenta (O9V), green (B1V), and yellow (B2V or later). White contours indicate the $^{13}$CO emission of component A (first contour level $\sim$ 8 rms) and the CO emission of feature 1 (first contour level $\sim$13 rms) , feature 2 (first contour level $\sim$12 rms), and feature 3 (first contour level $\sim$10 rms).   }
         \label{ysos-exc}
   \end{figure*}
To this aim, MSX sources were selected satisfying a flux quality Q $>$1 in the four bands. WISE sources with photometric flux uncertainties $>$ 0.2 mag and signal-to-noise ratio (S/N) $<$ 7 in the W1, W2, W3, and W4 bands, were rejected. Finally, we kept {\it Spitzer} sources with photometric uncertainties $<$ 0.2 mag in the four IRAC bands. Then, within a 14\arcmin\ box centered at  \radec = (12\h 35\m 44\s, $-$63\gra00\arcmin08\arcsec), we found a total of  2 MSX, 400 WISE, and 810 {\it Spitzer} sources fulfilling the selection criteria mentioned above. To identify candidate YSOs  we adopted the classification scheme described in \ \citet{lum02}, \citet{koe12}, and \citet{gut09} for the MSX, WISE, and {\it Spitzer}  sources, respectively. Several sources were found to qualify the above criteria, which are listed in Table \ref{tysos}. The table presents the designation of the sources, their coordinates, flux densities, information concerning the type of the protostellar object, matching with another source in the table, and coincidence with a  molecular or IR structure.

Among the MSX sources, we found one MYSO candidate, which coincides with the center of IRAS\,12326-6245, and one CHII region candidate, which coincides with a WISE Class II candidate (see below). Before attempting to identify the candidate YSOs from the listed WISE and {\it Spitzer} sources, we selected the non-YSO  sources with excess infrared emission, such as PAH emitting galaxies, broad line active galactic nuclei (AGNs), unresolved knots of shock emission, and PAH emission features.  Then, a total of 89 and 51 WISE and {\it Spitzer} sources, respectively, were dropped from the lists.
Among the remaining 311 WISE and 759 {\it Spitzer} sources, 12 (5 WISE and 7 {\it Spitzer}) were identified as Class I sources (i.e., sources with IR emission arising mainly from a dense
infalling envelope, including flat spectrum objects) and 60 (43 WISE and 17 {\it Spitzer}) as Class II sources (i.e., preMS stars with optically thick disks).

\begin{table*}[hbt!]
\caption{YSO candidates obtained from MSX, WISE, and {\it Spitzer} catalogs. \label{tysos}}
\begin{tabular}{c c c c c c c c l}
\hline\hline
\multicolumn{9}{c}{MSX sources} \\
 \hline
 $\#$ &  Designation &  $\alpha$ & $\delta$           &  F$_{8}$    &  F$_{12}$  &  F$_{14}$  &  F$_{21}$  & Notes \\
      &              &   [h m s] & [$\circ$ $\prime$ $\prime\prime$] &    [Jy]     &   [Jy]     &  [Jy]          &   [Jy]         &  (class, match with other YSO and      \\
      &&&&&&&&  with  $^{13}$CO/CO or IR structure)\\
 \hline
1 & G301.1364-00.2249  & 12 35 35.2 & --63 02 30 & 1.41  &   5.30  &   17.92  &   120.23  & (MYSO)\ \ MC3, WISE~$\#$6 \\
2 & G301.1300-00.2354 & 12 35 31.5 & --63 03 06 & 2.216  &  2.702  & 1.477  &  8.930 &  (CHII)\ \ MC3, WISE~$\#$3  \\
\hline
\multicolumn{9}{c}{ WISE sources} \\
\hline
$\#$ &  Designation &  $\alpha$ & $\delta$                          &     W1   &    W2    &  W3   &  W4   &  \\
     &              &   [h m s] & [$\circ$ $\prime$ $\prime\prime$] &    [mag]     &   [mag]     &  [mag]          &   [mag]         &        \\
\hline
3 &J123531.48-630306.4  &       12 35 31.5      &       --63 03 06.4    &       8.759 &       7.615&  3.207&  -0.535 &  (Class I), MC3, MSX~$\#$2 \\
4 &J123529.88-630417.1  &       12 35 29.9      &       --63 04 17.1    &       11.847& 10.812& 5.958&  2.858 &  (Class I), MC2 \\
5 &J123553.78-630012.6  &       12 35 53.8      &       --63 00 12.6    &       9.87&   8.769&  5.207&  1.367 & (Class I), (+)   \\
6 &J123535.13-630230.3  &       12 35 35.1      &       --63 02 30.3    &       9.49&   6.629&  1.011&  -3.604 &  (Class I), MC3, MSX~$\#$1 \\
7 &J123534.52-630147.8  &       12 35 34.5      &       --63 01 47.8    &       11.804  &10.715&        5.25&   2.989 & (Class I), MC3, (*)  \\
8 &J123529.71-630425.9  &       12 35 29.7      &       --63 04 25.9    &       10.829  &       9.987   &       6.386   &       4.376   &    (Class II), MC2 \\
9 &J123528.94-630432.6  &       12 35 28.9      &       --63 04 32.6    &       10.492  &       9.912   &       6.113   &       3.56    & (Class II), MC2\\
10 &J123531.19-630328.3 &       12 35 31.2      &       --63 03 28.3    &       10.324  &       9.984   &       6.269   &       3.648   &  (Class II), MC3\\
11 &J123538.44-630318.5 &       12 35 38.5      &       --63 03 18.5    &       13.115  &       12.557  &       6.902   &       3.138   & (Class II), MC3  \\
12 &J123537.22-630300.1 &       12      35      37.2    &       --63    03      00.1    &       11.906  &       10.999  &       8.069   &       1.563   & (Class II), MC3  \\
13& J123534.55-630311.1 &       12      35      34.6    &       --63    03      11.1    &       11.405  &       10.717  &       5.906   &       3.999   & (Class II), MC3\\
14 & J123628.80-630500.5 &      12      36      28.8    &       --63    05      00.5    &       10.379  &       9.698   &       6.464   &       4.797   & (Class II), Feature 2, {\it Spitzer}~$\#$61   \\
15 & J123606.78-630458.5 &      12      36      06.8    &       --63    04      58.5    &       10.812  &       10.396  &       9.114   &       6.107   & (Class II)\ \   \\
16 &  J123553.95-625717.3 & 12  35  53.9    &  - -62    57  17.3    &   12.834  &   12.427  &   5.898   &   3.722   & (Class II) \\
17 & J123556.74-625804.3&       12      35      56.7    &       --62    58      04.3    &       13.196  &       12.637  &       6.007   &       4.376   & (Class II)\   \\
18 & J123556.58-630058.1&       12      35      56.6    &       --63    00      58.1    &       12.681  &       12.309  &       6.877   &       5.598   & (Class II), MC5, (*) \\
19 & J123559.25-630043.5&       12      35      59.3    &       --63    00      43.5    &       12.784  &       12.065  &       6.416   &       4.826   & (Class II), MC5, (*) \\
20 & J123605.26-630000.2&       12      36      05.3    &       --63    00      00.2    &       12.44   &       11.992  &       7.689   &       5.894   & (Class II), MC5, (*)\\
21 & J123608.20-625952.4&       12      36      08.2    &       --62    59      52.4    &       13.621  &       12.973  &       7.051   &       5.06    & (Class II), MC5, (*)\\
22 & J123544.39-630151.3&       12      35      44.4    &       --63    01      51.3    &       12.362  &       11.598  &       7.181   &       4.462   & (Class II), MC3, (*) \\
23 & J123535.61-630129.4&       12      35      35.6    &       --63    01      29.4    &       11.64   &       11.000  &      5.101   &       3.019   & (Class II), MC3, (+)  \\
24 & J123533.44-630130.5&       12      35      33.4    &       --63    01      30.5    &       11.504  &       10.981  &       5.308   &       2.927   & (Class II), MC3, (+)\\
25 & J123539.82-630127.8&       12      35      39.8    &       --63    01      27.8    &       11.651  &       11.14   &       6.441   &       3.786   & (Class II), MC3, (+)  \\
26 & J123542.44-630148.3&   12 35 42.3      &   --63    01  48.3    &   11.042  &   10.237  &   6.363   & 2.654     & (Class II), MC3, (*), {\it Spitzer}~$\#$67 \\
27 & J123538.39-630139.4&       12      35      38.4    &       --63    01      39.4    &       10.882  &       10.354  &       5.752   &       2.963   & (Class II), MC3, (+)  \\
28 & J123525.27-630114.3&       12      35      25.3    &       --63    01      14.3    &       9.593   &       8.952   &       4.275   &       2.262   & (Class II), MC3, (+) \\
29 & J123528.77-630113.8&       12      35      28.8    &       --63    01      13.8    &       10.472  &       9.535   &       4.4         &    2.781   & (Class II), MC3, (+) \\
30 & J123522.97-630048.0&       12      35      23.0    &       --63    00      48.0    &       11.545  &       10.94   &       6.364   &       1.844   & (Class II), (+)  \\
31 & J123526.73-630156.8&       12      35      26.7    &       --63    01      56.8    &       11.679  &       10.821  &       5.627   &       2.865   & (Class II), MC3  \\
32 & J123521.24-630158.0&       12      35      21.2    &       --63    01      58.0    &       11.983  &       11.316  &       5.168   &       2.302   & (Class II), MC3, (+) \\
33 & J123525.41-630204.4&       12      35      25.4    &       --63    02      04.4    &       11.457  &       10.729  &       5.582   &       3.669   &(Class II), MC3    \\
34 & J123524.65-630218.0&       12      35      24.7    &       --63    02      18.0    &       11.548  &       10.785  &       5.839   &       3.296   & (Class II), MC3  \\
35 & J123525.28-630126.8&       12      35      25.3    &       --63    01      26.8    &       10.764  &       10.072  &       5.26    &       4.748   & (Class II), MC3, (+)  \\
36 & J123552.92-625938.6&       12      35      52.9    &       --62    59      38.6    &       13.013  &       12.391  &       6.458   &       4.429   & (Class II), (+)  \\
37 & J123546.50-630010.4&       12      35      46.5    &       --63    00      10.4    &       13.483  &       12.56   &       6.118   &       4.25    & (Class II), (+) \\
38 & J123551.36-630013.5&       12      35      51.4    &       --63    00      13.5    &       12.246  &       11.293  &       5.803   &       4.145   & (Class II), (+) \\
39 & J123550.35-630001.0&       12      35      50.4    &       --63    00      01.0    &       12.141  &       11.6    &       6.098   &       4.645   & (Class II), (+) \\
40 & J123541.99-630100.8&       12      35      42.0    &       --63    01      00.8    &       11.681  &       10.913  &       4.941   &       3.329   & (Class II), (+) \\
41 & J123543.50-630002.3&       12      35      43.5    &       --63    00      02.3    &       13.039  &       12.656  &       7.897   &       4.866   & (Class II), (+)  \\
42 & J123543.44-630013.6&       12      35      43.4    &       --63    00      13.6    &       12.344  &       12.003  &       6.657   &       4.019   & (Class II), (+) \\
43 & J123542.11-630037.9&       12      35      42.1    &       --63    00      37.9    &       12.257  &       11.573  &       5.277   &       3.368   & (Class II), (+) \\
44 & J123543.69-630050.6&       12      35      43.7    &       --63    00      50.6    &       11.771  &       11.174  &       5.177   &       3.43    & (Class II), (+) \\
45 & J123545.43-630023.2&       12      35      45.4    &       --63    00      23.2    &       11.848  &       11.265  &       5.344   &       3.169   & (Class II), (+) \\
46 & J123544.72-630038.4&       12      35      44.7    &       --63    00      38.4    &       12.206  &       11.608  &       5.446   &       4.213   & (Class II), (+) \\
47 & J123543.60-630027.6&       12      35      43.6    &       --63    00      27.6    &       12.647  &       12.25   &       6.496   &       5.909   & (Class II), (+) \\
48 & J123544.53-625720.1&       12      35      44.5    &       --62    57      20.1    &       7.801   &       6.65    &       5.019   &       4.336   & (Class II), {\it Spitzer}~$\#$52  \\
49 &  J123454.60-630009.5&      12      34      54.6    &       --63    00      09.5    &       11.662  &       11.15   &       8.66    &       6.146   & (Class II), Feature 3, {\it Spitzer}~$\#$72\\
50 & J123558.47-625603.9&       12      35      58.5    &       --62    56      03.9    &       10.983  &       10.688  &       7.725   &       5.376   & (Class II)\ \   \\
\hline
 
\end{tabular}
\end{table*}

\addtocounter{table}{-1}

\begin{table*}[ht]
\caption{ Continued. }
\begin{tabular}{c c c c c c c c l}
\hline\hline
\multicolumn{9}{c}{ {\it Spitzer} sources} \\
\hline
$\#$ &  Designation &  $\alpha$  & $\delta$   & 3.6 $\mu$m   &   4.5 $\mu$m    &   5.8 $\mu$m   &    8.0 $\mu$m &  \\
     &              &   [h m s] & [$\circ$ $\prime$ $\prime\prime$] &    [mag]     &   [mag]     &  [mag]          &   [mag]         &   \\
\hline
51   &  G301.1340-00.0894       &  12   35      38.3    &   --62  54 21.5         & 12.44        &       11.532  &       10.577  & 9.693 & (Class I) \ \ \\
52   &  G301.1489-00.1383       &  12   35      44.5    &   --62  57 20.5         & 8.62         & 7.55          & 6.595 & 5.873 &  (Class I), WISE~$\#$48  \\
53   &  G301.1866-00.1701       &  12   36      03.4    &   --62  59 23.0         & 12.424       &       11.565  &       10.773  &       9.521   &(Class I) \ \  \\
54  &  G301.1955-00.2012        &  12   36      07.1    &   --63  01 16.6   &     11.595          & 10.529        &       9.709 & 8.791   & (Class I), MC5, (*) \\
55  &  G301.1766-00.2110        &  12   35      56.8    &   --63  01 47.8  & 14.545               & 12.909                & 11.945        &       12.002  & (Class I), MC4 \\
56  & G301.2000-00.1534    &  12        36      11.0    &   --62  58 25.8         & 13.885       &       12.788          & 11.921        &       10.916  & (Class I) \ \   \\
57   &  G301.1338-00.2711       &  12   35      32.2    &       --63  05 14.4      &  12.721     &  11.54&       10.612  &  9.959& (Class I), MC1 \\
58   &  G301.1402-00.2298 &  12 35      37.0  &         --63  02 47.4     &  12.521      &  11.843 &     11.118  &  10.65&  (Class II), MC3 \\
59   &  G301.1395-00.2303       &  12   35      36.6    &   --63  02 49.0   &  11.273&    10.604 &        10.012  &  9.482 & (Class II), MC3  \\
60   &  G301.1523-00.2264       &  12   35      43.5    &   --63  02 37.8         &  12.498      &  12.146       &  11.893       &  10.887 & (Class II), MC3  \\
61   &  G301.2400-00.2609 &  12 36      28.8    &   --63  05 00.6         &      10.204  &       9.729   &       9.231   &       7.869   & (Class II), Feature 2, WISE~$\#$14  \\
62   &  G301.1973-00.2637 &  12 36      06.1 &          --63  05 01.6     & 11.215       &       10.808          & 10.571        &       10.131  & (Class II)\ \  \\
63   &  G301.2169-00.2211       &  12   36      17.8  &         --63  02 32.7   &        13.615&         12.915  &  12.217&      11.599&  (Class II), MC6   \\
64  &  G301.1745-00.2102        &  12   35      55.7    &   --63  01 44.5         &      13.464  &       12.670  &       12.148          & 10.998        & (Class II)\ \   \\
65   &  G301.2153-00.1838       &  12   36      18.1    &   --63  00 18.3         & 7.683        &       7.482   &       7.007           & 6.772 & (Class II), MC5, (*)    \\
66   &  G301.1311-00.2271       &  12   35      32.3    &   --63  02 35.7         & 12.992       &       11.898  &       11.354 &        10.504  & (Class II), MC3 \\
67   &  G301.1500-00.2124       &  12   35      42.7    &   --63  01 47.0         & 11.332               & 10.95 &       10.353 &        9.342   & (Class II), MC3, (*), WISE~$\#$26 \\
68   &  G301.1402-00.2155       &  12   35      37.5    &   --63  01 56.0         & 11.273       &       10.637  &       10.065 &        9.522   & (Class II), MC3, (*)  \\
69   &  G301.1432-00.2154       &  12   35      39.9    &   --63  01 56.3   & 12.104      &       11.591  &       11.068 &        10.594  & (Class II), MC3, (*) \\
70   &  G301.1090-00.1906       &  12   35      21.8    &   --63  00 19.6         & 13.625       &       12.923  &       12.308 &        11.642  & (Class II)\ \   \\
71   &  G301.1250-00.2201       &  12   35      29.3    &   --63  02 09.2         & 10.166       &       9.538   &       8.947           & 8.425 & (Class II), MC3 \\
72   &  G301.0569-00.1911       &  12   34      54.3    &   --63  00 09.5         & 12.334       &       11.901  &       11.645          & 10.713 & (Class II), Feature 3  \\
73   &  G301.0578-00.1910       &  12   34      54.8    &   --63  00 09.4         & 12.095       &       11.671  &       11.336          & 10.556        & (Class II), Feature 3, WISE~$\#$49   \\
74   &  G301.1525-00.0712       &  12   35      48.6    &   --62  53 20.2         & 12.940       & 12.224 & 11.714& 11.15& (Class II)\ \ \\
\hline
\end{tabular} 
\end{table*}
 
 The location of the identified candidate YSOs is shown in Fig.~\ref{ysos-exc}, where the 8.0 and 5.8 $\mu$m  emissions together with the emission  of both  the  $^{13}$CO and CO structures  (i.e, component A and features 1, 2, and 3) are also displayed.  As expected, the bulk of the candidate YSOs are seen projected onto the densest molecular gas (component A). A noticeable feature is the concentration of sources  along the two brightest  PDR areas which, as mentioned in Sect. \ref{compa}, delineates the borders of the molecular condensations. Namely,  the WISE sources $\#$5, $\#$23, $\#$24, $\#$25, $\#$27, $\#$28, $\#$29, $\#$30, $\#$32, and  $\#$35 to  $\#$47, seem to be related to the PDR closer to the ionized gas bordering the molecular emission of MC3 and MC4 (noted as (+) in the last column of Table~\ref{tysos}), while WISE sources $\#$7, $\#$18, $\#$19, $\#$20, $\#$21, $\#$22, and $\#$26, and {\it Spitzer} sources $\#$54, $\#$65, $\#$67, $\#$68,  and $\#$69  lie along the PDR    that lies farther away from the ionized gas and related to the densest parts of MC3, MC4, and MC5 (noted as (*) in Table~\ref{tysos}).  Not surprisingly, another  appreciable concentration of candidate YSOs, namely WISE sources $\#$3, $\#$6, $\#$10,   $\#$11,   $\#$12,   and $\#$13,   {\it Spitzer} sources   $\#$58,   $\#$59,   $\#$60,   $\#$66, and    $\#$71, and MSX sources $\#$1 and $\#$2  can be noticed in the region of the molecular concentration MC3.

  \section{Discussion}\label{disc}
 
\subsection{A model for S169 and its molecular environment}\label{model}

Since  IR bubbles are believed to be born  within dense molecular clouds,  the classical scenario predicts that, in a uniform medium,  the molecular gas around the bubble should expand spherically. Under that assumption, a molecular shell with a central velocity V$_0$ and an expansion velocity V$_{\rm exp}$ should depict in a position-position diagram a "disk-ring" pattern when observed at different velocities. At V$_0$ (corresponding to the systemic velocity of the bubble), the shell should attain its maximum diameter; whereas, at extreme velocities (either negative or positive with respect to V$_0$), the molecular emission should resemble a disk. At intermediate velocities,  the radius of the ring shrinks as the extreme velocity is approached.  Although these features were never observed   all together in the molecular gas around IR bubbles,  the behavior of the molecular gas around S169 certainly clashes with  that of the classical IR bubble model. 

In order to address  the observed characteristics of the molecular gas associated with S169, we present  a simple model taking into account its morphology, relative position with respect to the IR and H$\alpha$ emissions  and the velocity  intervals as constraints. Needless to say, we assume that component A, feature 1, feature 2, and feature 3 are physically associated with S169 --  a well-grounded conclusion given their good morphological correspondence with the IR nebula.
\begin{figure}
   \centering
   \includegraphics[width=8cm]{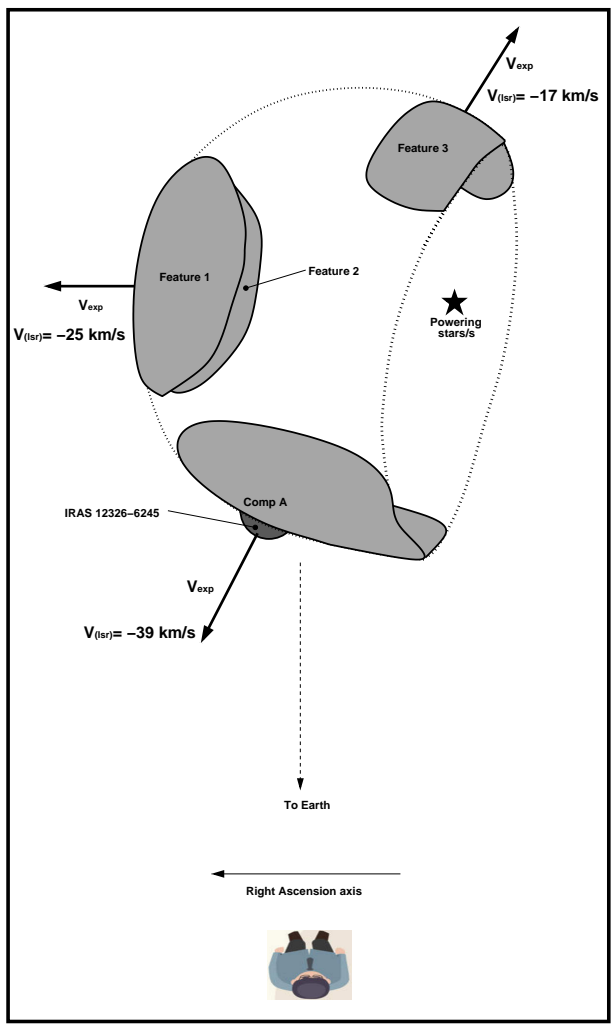}
      \caption{ Simple sketch of the proposed model   to  explain the morphology and velocity of the molecular gas associated with S169. The right ascension axis, as visualized by the observer, is indicated as reference (the  declination axis is perpendicular to the plane of the sketch, increasing towards the reader).    }
         \label{modelo}
   \end{figure}
Hence, we propose that component A and  features 1, 2, and 3 could be explained in terms of an expanding, partially complete semispherical structure. A simple sketch is presented in Fig.~\ref{modelo}, where the right ascension axis  is indicated for the sake of clarity (the  declination axis is then perpendicular to the sketch, increasing towards the reader). 
The expansion of the molecular gas is then revealed by the velocity of the different molecular structures. The molecular gas approaching in the direction of Earth is depicted by the molecular structure at more negative velocity (component A) with an approaching velocity of V$_{\rm appr}$ $\approx$ $-$39 \kms, while the molecular gas receding is depicted by the structure at more positive velocity (feature 3)  with a receding velocity of V$_{\rm rec}$ $\approx$  $-$17 \kms. Molecular structures at intermediate velocities (features 1 and 2) thus represent the molecular gas  at V$_0$ $\approx$ $-$25 \kms\   expanding approximately in perpendicular directions with respect to the direction of  Earth. A glance at Fig.~\ref{combyc} suggests that the expansion velocity of features 1 and 2 might have components in the direction of the declination axis, which is positive for the case of feature 1 and negative for the case of feature 2 (this cannot be depicted in the two-dimension sketch  presented in  Fig.~\ref{modelo}). 

Therefore, with this simple model we are able to explain not only the morphology and kinematics of the molecular gas, but also the characteristics of the ionized gas (traced by the radio continuum emission), which seems to be expanding against the molecular gas towards the east, north, and south, while  it is expanding freely to the west, probably in a sort of a champagne-flow effect.

 According to the model, a rough estimation of the expansion velocity of the molecular gas around S169 may be simply obtained by: 
\begin{equation}
 \quad  {\rm V}_{\rm exp} = \frac{|{\rm V}_{\rm appr}| - |{\rm V}_{\rm rec}|}{2 \times\ {\rm cos}(\theta)} 
\label{vexp}
,\end{equation} 
with $\theta$ as the angle between the line of sight and the direction of V$_{\rm exp}$ of component A and feature 2. Based on the morphology of the IR nebula, a conservative value $\theta$=25\gra\ can be adopted, and   the obtained expansion velocity turns out to be V$_{\rm exp}$ $\sim$ 12 \kms.   Having the expansion velocity, the dynamical age of the nebula, $t_{\rm dyn}$, can be derived considering a radius  of 2.3$\pm$0.8  pc  ($\sim$ 4$'$ at a distance of 2.03$^{+0.77}_{-0.61}$ kpc), which turns out to be in the range of (1.2 - 2.6) $\times$ 10$^5$ yr.

\subsection{Distance to S169 and IRAS\,12326-6245}\label{dist}

As noted in Sect.~1, the adopted distance in the literature for IRAS\,12326-6245 is 4.4 kpc. This value  was first estimated by  \citet{ost97}, who performed a study of a number of southern IRAS sources using IR continuum and millimeter line data. The authors adopted distances  that were taken from the literature where possible (either directly or by kinematic and spatial association with another source of known distance). Otherwise, kinematical  distances were determined using the central velocity of the CS(2-1) or CO(2-1) emission lines and the Galactic rotation curve of \citet{cle85}. For the case of IRAS\,12326-6245, a kinematical distance could not be determined by the authors since the central  velocity of the CS and CO lines ($-$39.4 \kms\ and $-$39.7 \kms, respectively) are forbidden according to  the Galactic rotation model. Then the authors adopted for  IRAS\,12326-6245  the distance to the source IRAS\,1283-6128 (4.4 kpc), which is kinematically close to within 0.5 \kms \ as well as spatially to within 2\gra. 
\begin{figure}
  \includegraphics[width=9cm]{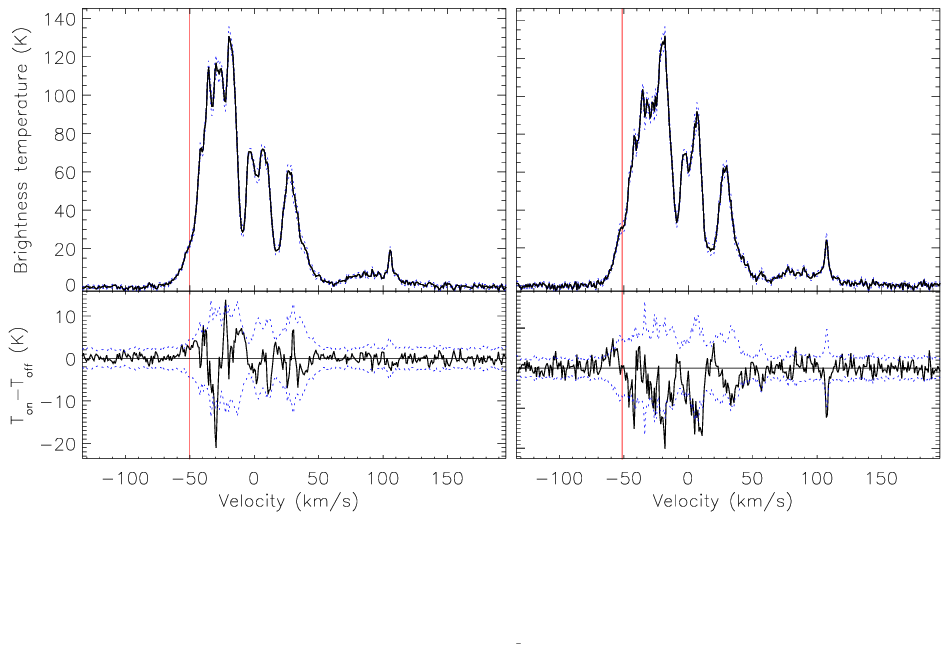}
  \caption{{\it Left panel.} HI emission/absorption profile towards the bright southern component of the radio continuum arc-shaped feature. {\it Right panel.} Same for the extragalactic source at the north of the nebula. The error envelopes are plotted with blue dotted lines. Vertical red lines indicate the velocity of the tangent point. }
\label{perfs}
\end{figure}

In Sect.~3.1, we offer a confirmation of  the physical association between S169 and IRAS\,12326-6245. Furthermore, according to the model proposed in Sect.~\ref{model}, the source IRAS\,12326-6245 (embedded in component A) is expanding at a velocity coarsely estimated as V$_{\rm exp}$ $\sim$ 12 \kms\  with respect to the systemic velocity of the bubble (V$_0\approx-$25 \kms). Assuming that our model is well-suited to explain the characteristics of the molecular gas around the bubble nebula, this leads to the unavoidable conclusion that the velocity reported in the literature for IRAS\,12326-6245 ($\sim$ $-$39 \kms) does not represent  the velocity of the ISM in the surroundings of both, the IRAS source and S169. This is understandable considering that previous studies of IRAS\,12326-6245 were not performed in the whole context of the nebula S169, but rather focused only on the IRAS source.  Instead, we believe that the velocity of features 1 and 2 ($\sim$ $-$25 \kms), corresponding approximately to the systemic velocity of the bubble nebula, is more suitable for determining   the kinematical distance for both, S169 and  IRAS\,12326-6245. Furthermore, in a study of the rotation curve in the Southern Galaxy, \citet{alv90} found  an excess anomally of $+$12.2 \kms\ in the terminal velocities in the Galactic longitude range between $l$ = 280 and $l$ = 312, with respect to the trend in the rest of the IV Galactic quadrant. This excess includes the region studied here since we have our source at  Galactic coordinates  $(l,b)$ = (301.134, --0.225). Next, by adopting  a velocity of $-$25 \kms, and using  the rotation curve of  \citet{reid14} with the Monte Carlo method\footnote{https://www.treywenger.com/kd/index.php} (see \citealt{weng18} for details) and a conservative velocity uncertainty of 1 \kms,  two kinematical distances are obtained, namely: a near kinematical distance of 2.03$^{+0.77}_{-0.61}$ kpc, and a far kinematical distance of 6.59$^{+0.62}_{-0.77}$  kpc.

To resolve the twofold ambiguity in kinematic distance, we made use of the \hi \ data from the SGPS constructing absorption profiles towards the bright southern component of the radio continuum arc-shaped feature and, for comparison, towards the extragalactic source (see Sect.~\ref{ioniz}). The radio continuum data were convolved to a beam of 2\arcmin\  to make them compatible with the resolution of the \hi \ data. The expected off-source profile was computed using a bilinear interpolation method as described in \citet{reynoso2014}. The results are shown in Fig.~\ref{perfs}. From the emission spectra, the tangent point lies at $\sim -50$ km s$^{-1}$ (indicated with red vertical lines in Fig.~\ref{perfs}), while emission at the outer Galaxy is observed as far as $\sim +110$ km s$^{-1}$. The extragalactic source (right panel) exhibits, as expected, absorption features up to $\sim -45$ km s$^{-1}$, coincident with the tangent point within a few km s$^{-1}$, a difference that can be ascribed to the ISM turbulence. Another clear absorption feature at $\sim$ +110 km s$^{-1}$ confirms the extragalactic nature of this source.  In contrast, the profile corresponding to the arc-shaped feature (left panel) displays the last significant absorption feature at about $\sim -30$ km s$^{-1}$. Considering the absence of absorption in a gap of about 20 km s$^{-1}$ between this velocity and the tangent point, it is very unlikely that S169 is located at the far side of the Solar circle.  Based on these results, we conclude that the near kinematical distance (2.03$^{+0.77}_{-0.61}$ kpc) is the most adequate for S169. The detection of H$\alpha$ emission at the center of the nebula (see Fig.\ref{irac-sc}), which is likely the optical counterpart of the ionized nebula, gives more support to this conclusion since the  absorption that could be produced by the ISM along the far kinematical distance ($\sim$ 6.6 kpc) would make the H$\alpha$ emission much fainter or even undetectable. Furthermore, the whole molecular cloud that contains IRAS\,12326-6245  has been identified to be in the near side of the Carina Spiral Arm by \citet{coh85}.
 
\subsection{Possible triggered star formation}\label{trig}

  The spatial distribution of the candidate YSOs along the borders of the IR nebula and the densest part of the molecular gas (see Sect.~\ref{ysos}) is indicative of a 
triggered star-forming process such as collect-and-collapse (C$\&$C; \citealt{elme77}) or radiative driven implosion (RDI; \citealt{lefl94})   might be acting on the nebula.

In order to test the C$\&$C mechanism, we use the classical model of \citet{wi94} for expanding \hii\ regions. Then, we estimated the fragmentation time, $t_{\rm frag}$, and the radius  of the \hii\ region when the fragmentation occurs, $R_{\rm frag}$, as: 
\begin{equation}  
\quad t_{\rm frag}\ \  =\ 1.56\  {a_{0.2}}^{7/11}\   n_3^{-5/11}\  (N_{49})^{-1/11}   \ \ 10^6 \ \ {\rm yr}
\label{tfrag}                       
,\end{equation} 
\begin{equation}  
\quad R_{\rm frag}\ \ =\ 5.8\ {a_{0.2}}^{4/11}\ n_3^{-6/11}\   (N_{49})^{1/11}  \ \ {\rm pc}
\label{rfrag}                       
,\end{equation}
with  $a_{0.2}$  as the isothermal sound speed in the compressed layer in units of 0.2 \kms\ ($a_s/0.2$ \kms),  $n_3$ the  surrounding homogeneous infinite medium into which the \hii\ region expands, in units of 10$^3$ cm$^{-3}$ ($n_0/10^3$ cm$^{-3}$),  and  $N_{49}$  the number of ionizing Lyman continuum photons, in units of 10$^{49}$\,s$^{-1}$ ($N^*_{\rm Lyc}$/10$^{49}$ s$^{-1}$).  We adopt $a_{s}$ =  0.2 \kms\ for the collected layer (likely a lower limit  since both turbulence and extra heating from intense subLyman continuum photons leaking from the \hii\ region could increase this value)   and $N^*_{\rm Lyc}$ =  2.5 $\times$ 10$^{47}$ s$^{-1}$  (see Sect.~\ref{ioniz}). To estimate the initial density, $n_0$,  we averaged the total mass of the molecular gas around S169 \hbox{($\sim$ (7.4 $\pm$ 4.4) $\times$ 10$^3$ \msun)} over a sphere of 2.3 $\pm$ 0.8  pc  in radius, which yields  a density in the range of    $n_0$ =  (0.35 - 12) $\times$ 10$^3$ cm$^{-3}$. Thus,  we obtain  $t_{\rm frag}$ in the range between 0.5 and 2.2 Myr  and $R_{\rm frag}$ between  1.1 and 7.4 pc.  
 
A comparison between $R_{\rm frag}$ and the present radius of the nebula suggests (within errors) that the  C$\&$C process may be acting in S169. On the contrary, the estimated fragmentation time  is higher than the  $t_{\rm dyn}$ derived for the region in Sect.~\ref{model}. Thus, there is no conclusive evidence to support that  protostellar objects  at the borders of S169 could have been formed as a  result of the C$\&$C mechanism. We ought to keep in mind, however,  that the model of \citet{wi94} is valid for evolution of the \hii\ regions only in a uniform molecular environment, which is certainly not  the case with S169. 

 Regarding the possibility that the RDI process is taking place  in certain regions of the nebula, a more detailed study is required, with better angular resolutions of both the molecular and radio continuum data. Nevertheless, we can clearly conclude that the presence of the intense PDR emission detected as a bright rim of the densest molecular feature suggests that the RDI process could be responsible for  the formation of  some candidate YSOs.   In fact, an alternative scenario for the star formation in S\,169 could be that proposed by \cite{wal15}, which considers the action of massive stars on a fractal molecular cloud resulting in a hybrid mechanism that combines elements of C\&C and RDI and which has the capability to generate shell structures with dense clumps within them. If this were the case for S169, the denser region where IRAS\,12326-6245 is located could be explained by considering an initial non-uniform molecular cloud with a significant increase in density and, thus, the RDI process  could be the main action responsible for  the formation of the candidate YSOs located behind this border and, in particular, for the IR source.

\section{Summary}\label{concl}

As  part  of  a  broader project  aimed  at  characterizing and studying  the  physical  properties  of  Galactic IR  bubbles  and their surroundings, we present a multiwavelength analysis of the southern IR dust bubble S169 associated with the massive star forming core IRAS\,12326-6245. To analyze the characteristics of the molecular gas in the whole nebula, we use  \coa, \cob, \coc \  line data, while for the IRAS source we also use \hcn, and \hco\ data, all   obtained  with  the APEX telescope. To study the characteristics of the dust and ionized gas, as well as to investigate the presence of stellar and protostellar objects in the nebula, we use archival data.

 The analysis of the \coa, \cob, and \coc\ data allows  us to identify the molecular gas linked to the nebula. We report three molecular components at $\sim$ $-$39, $-$25, and $-$17 \kms\ (components A, B, and C), which are morphologically correlated with different regions of the IR nebula. The component at $-$39 \kms\ (component A) is associated with the brightest part of the IR nebula (at the southern region) and is the only one detected  in \cob\ and \coc\ emissions. For this component, we have identified six molecular condensations (namely MC1, MC2, MC3, MC4, MC5, and  MC6) based on  the $^{13}$CO and C$^{18}$O emissions. These condensations have masses between $\sim$ 90 and 5500 \msun, and H$_2$ column densities between $\sim$ 6$\times$10$^{21}$ and 1$\times$10$^{23}$ cm$^{-2}$. The densest one (MC3) is the molecular counterpart of IRAS\,12326-6245, which demonstrates the physical association between the nebula and the IRAS source.
 An LTE analysis of the \hco\ and \hcn\ lines on this source, assuming 50 and 150 K, respectively, indicates column densities of $N$(HCO$^+$) = (5.2 $\pm$ 0.1) $\times$ 10$^{13}$~cm$^{-2}$ and $N$(HCN) = (1.9 $\pm$ 0.5) $\times$ 10$^{14}$~cm$^{-2}$. For this condensation, an H$_2$ column density up to $\sim$ 8$\times$10$^{23}$ cm$^{-2}$ is obtained from the emission at 870 $\mu$m.
     
The molecular components at $-$25 and $-$17 \kms\ (components B and C) seem to be associated with the faintest and more external regions of the IR nebula. Unlike component A, components B and C  are not detected in $^{13}$CO and C$^{18}$O emissions, which indicates that they are composed of low-density gas. This can be confirmed from {\it Herschel} images at 70, 160, and 350  $\mu$m.

With the purpose of explaining the spatial distribution and velocity of components A, B, and C, we propose a very simple model, which consists of a partially complete semisphere structure expanding at $\sim$ 12 \kms. According to the model, component A represents the molecular gas approaching at a velocity of $-$39 \kms, \  while component C (feature 3) the molecular gas receding at a velocity of $-$17 \kms. The systemic velocity of the molecular gas associated with the bubble is then $-$25 \kms, which is the velocity of component B (features 1 and 2). The  distribution in  the radio continuum emission at 843 MHz suggests an \hii\ region   bounded by ionization to the east, north and south,  and by density to the west. This appears to be  in line with the model proposed for the molecular gas.  
The model proposed for the molecular gas brings an additional discussion about the distance of  IRAS\,12326-6245 (and consequently S169) since the radial velocity adopted in the literature for the IRAS source ($\sim$ --39 \kms) would not be representative of the velocity of the ISM in its surroundings. Instead, we believe that the systemic velocity of the bubble (--25 \kms) is more adequate for determining its kinematical distance. Then, using the Galactic rotation model and HI absorption profiles,  we determine a kinematical distance of 2.03$^{+0.77}_{-0.61}$ kpc for S169 and  IRAS\,12326-6245.

Using point source catalogs, we identify10  ionizing stars candidates   projected  onto the cavity. They have spectral types between O9V and B2, which are necessary to sustain the current level of ionization of the \hii\ region. We also identify a number of candidate YSOs projected mostly onto \hbox{component A}, more precisely: MC3, MC4, and MC5,   confirming that active star formation has developed along the borders of the bubble. After  comparing the fragmentation time and fragmentation radius with the age and current radius of the nebula, we  cannot assert that the C$\&$C process is acting in the collected layers of gas at the edge of the bubble. We do keep in mind, however, the limitations of the models applied to S169 and the possibility that other triggering star-forming processes, such as RDI or a combination of both, RDI and C$\&$C,  could be acting in the region.

In summary, the infrared bubble S169  is a relatively young HII region that has profoundly affected its surroundings, creating a molecular shell that continues to possess a large expanding motion and where several condensations host candidate protostellar objects in their dense interiors, with the source IRAS\,12326-6245 standing out as the most striking among them.

\begin{acknowledgements}

We would like to thank the anonymous referee for his/her helpful comments and suggestions that led to the improvement  of this paper. N.U.D and M.A.C acknowledge support from UNLP, PPID G005 and CONICET grant PIP 112$-$201701$-$00507 (Argentina). L.B. and R.F. acknowledges support from CONICYT project Basal AFB-170002. E.M. acknowledges support from the Brazilian agency CNPq (grant 150465/2019-0). L.A.S acknowledges support from UNLP  PPID G005 grant.  S.C., L.A.S. and E.M.R. are partially funded by CONICET grant PIP 112-201701-00604 (Argentina). 
\end{acknowledgements}
%
\bibliographystyle{aa} 
\bibliography{bibliografia.bib} 
%





\end{document}